\documentclass[pre,showpacs,showkeys]{revtex4}

\usepackage{amssymb}
\usepackage{amsmath}
\usepackage{graphicx}
\usepackage{amsbsy}
\usepackage{color}
\usepackage{bm}

\newcommand{\bq}{\begin{eqnarray}}
\newcommand{\eq}{\end{eqnarray}}
\newcommand{\bqn}{\begin{eqnarray*}}
\newcommand{\eqn}{\end{eqnarray*}}
\newcommand{\rr}{{\bf r}}
\newcommand{\RR}{{\bf R}}
\newcommand{\kk}{{\bf k}}
\newcommand{\bb}{{\bf b}}

\newcommand{\an}[1]{{#1}}
\newcommand{\bn}[1]{{#1}}

\begin{document}
\title{Field-theoretical approach to a dense polymer with an ideal
    binary mixture of clustering centers}

\author{Riccardo Fantoni}
\email{rfantoni@ts.infn.it}
\affiliation{National Institute for Theoretical Physics (NITheP) and
Institute of Theoretical Physics, Stellenbosch 7600, South Africa}
\author{Kristian K. M\"uller-Nedebock}
\email{kkmn@physics.sun.ac.za}
\affiliation{Institute of Theoretical Physics, Department of Physics,
Stellenbosch University, Private Bag X1, Matieland 7602, South Africa}

\date{\today}

\begin{abstract}
We propose a field-theoretical approach to a polymer system immersed 
in an ideal mixture of clustering centers.  The system contains several 
species of these clustering centers with different functionality, each 
of which connects a fixed number segments of the \an{chain} to each other.  
The field-theory is solved using the saddle point approximation and 
evaluated for dense polymer melts using the Random Phase Approximation.  
We find a short-ranged effective inter-segment interaction with strength
dependent on the average segment density and discuss the structure
factor within this approximation. We also determine the fractions of 
linkers of the different functionalities.
\end{abstract}


\maketitle
\section{Introduction}
\label{sec:introduction}

In this work we consider a system where \an{a single polymer chain is}
immersed in 
an ideal binary mixture of clustering centers.  The study of the
resulting associating polymers has a long history (see, e.g., in
\cite{HoyFredrickson2009,RubinsteinDobrynin1999} and references
in them). Models for the reversible gelation of polymers range from
the consideration of pairwise associations of sticky chain segments
\cite{SemenovRubinstein1998} to the formation of arbitrary size
clusters due to association of dipolar elements in polymer chains
\cite{Muthukumar1996}.  Whereas a large number of theoretical treatments
model association with some form of crosslinking, {\sl i.e.} linking of two
segments only, or arbitrary size clusters, we present a treatment for
a small number of species of clusters consisting of a fixed number of
polymer segments.  This falls into the so-called closed
multimerization scenario \cite{NyrkovaSemenov2005}.

The reversible or permanent linking of polymer chains or sections of 
polymers has been a topic of extensive investigations for many decades
in a wide range of systems.  The general statistical physical scenario
generally requires evaluating both the 
the statistical physics of the chains within a certain linked scenario
as well as a summation over all compatible modes of linking the polymer
constituents.  Independent of whether the linking (crosslinking,
aggregation, clustering, type of polycondensation, etc.)  
is permanent or reversible the topologically and geometrically
permissible combinations of linking or clustering need to be
evaluated, albeit with different strategies for quenched or annealed
situations~\cite{Deam1976}.  Therefore it is natural to think in terms
of the enumeration of graphs, as extensively reviewed by Kuchanov, et
al.~\cite{Kuchanov1988}. The ideas can then be applied to a variety of
systems, such as associating telechelics~\cite{Ermoshkin1999b},
polycondensation~\cite{Kuchanov2004}, polymers with multiply
aggregating groups~\cite{Kudlay2001} and general thermally reversible
aggregation, clustering or association
\cite{Sciortino2007,Kalyuzhnyi1998,Nagarajan1989,Ganazzoli1999,Tanaka1990,Xu1996,Semenov2009,Loverde2005,Erukhimovich1999}.   

There are different possibilities in which the scenario of fixed
functionality clustering can be realized.  One can think of the
segments of the chain connected to particles or sidechains that
assemble into structures with closed shells, akin to Janus particles,
where it was recently shown that monodisperse ten-particle micelles,
and forty-particle vesicles, are the thermodynamically dominant assembled
structures {\cite{Sciortino2009,Giacometti2009,Fantoni2011}}.  

We are interested in the properties of a solution of such \an{a
polymer} with clustering centers and in the relative dominance of
co-existing clusters with different, yet fixed, functionalities.  To
this end we reformulate a field-theoretical method originally proposed
by Edwards for permanent, arbitrary-functional end-linking of chains
\cite{Edwards1988}. The resultant field-theory is highly
non-linear, but can be treated analytically and numerically, offering
an additional theoretical tool to address questions on the formation
of localized, reversible structures of polymer chains. 

\bn{ As already mentioned, in order to compute the partition function
  or free energy one needs to evaluate the polymer chain conformations
  subject to the restrictions imposed by the functionalities of the
  linkers and include all possibilities of linking or cluster
  formation.  This is because we are modeling a strong type of
  aggregation with fixed functionalities, where the clusters are
  well-defined and local.  In other words, all permissible graphs must
  be generated, their connectivity restrictions be imposed on the
  polymer chains and weighted appropriately by Boltzmann factors.
  Kuchanov \emph{et al.}~\cite{Kuchanov1988}, in their exposition of
  strategies to do this, also point out the very clear analogy between
  enumeration of spatially embedded graphs and Feynman diagrams from
  field theories.  The obvious utility of a field-theoretical tactic
  lies in the large spectrum of available approximation techniques and
  graphical expansions but also in the freedom to choose the precise
  manner of implementing the additional fields.  }

\bn{ In this paper we introduce additional fields, with the associated
  functional integration, whose role is to produce the desirable
  linking, network-formation or aggregate possibilites as well as
  enforcing the spatial consequences of this on the appropriate
  monomers.  The current approach is similar to the those used in
  Refs.~\cite{Edwards1988,Kuchanov1988} and in work that can be seen
  as precursor to the current formalism
  \cite{Edwards1970a,Edwards1970b}.  Whereas the specific systems
  investigated in Refs.~\cite{Edwards1988} and \cite{Kuchanov1988} are
  addressed such that the ends of polymer chains or of star-like
  polymer units can associate, respectively, the system under
  investigation here deals with aggregation of segments of a polymer
  chain.  We show that this system allows a formulation of the field
  theory that has the mathematically advantageous property that it is
  \emph{local} in the introduced fields and these additional fields
  also couple to the density of monomers in a local way.  }

\bn{ Clearly the sum over aggregated states, by whatever method
  derived and approximated, would generally impose a complicated form
  onto the polymer conformational averages.  This is also the case for
  theories with additional fields following integration over the
  fields.  In approximating the functional integral over fields one
  expects to find nonlinear integral equations for these fields in the kinds of
  self-consistent field theory calculation that emerge in models for
  many systems (as in \cite{Kuchanov1988,Edwards1988}).  However, in
  our formulation the saddle point equations related to the
  additional fields (and taken before integration over the polymer
  degrees of freedom) turn out to be only algebraic, albeit nonlinear,
  providing a significant advantage in tractability in comparison to
  nonlinear integral equations obtained within other strategies (for
  other systems).  Analysis of the stability, etc. of the resulting
  theory is also relatively simpler.  Completing the integration over
  the fields incorporates the clustering into the remaining weight for
  the conformations of the chain (also in a local manner)
  giving the ``structural'' contribution that is taken together
  with the remaining polymer-polymer interactions.  }

\bn{ Yet other path integration techniques have made use of generating
  functional approaches to enumerate tree-like
  configurations~\cite{Gordon1962,Gordon1964,Mohan2010} in associating
  systems.  Motivated by a wide range of physical scenarios under
  which polymer chains can aggregate, many different methods (mainly
  not field-theoretical in the sense as here) have been utilized in
  determining the contribution of certain classes of connections,
  ranging from summations over a subset of looped
  conformations~\cite{Muthukumar1996}, sums of tree-like
  graphs~\cite{BohbotRaviv2004} and trees with cycles to analyses for
  stickers \cite{Ermoshkin1999b,SemenovRubinstein1998,Cates1986}.
 Typically the effective
  polymer-polymer contributions can then be dealt with through a
  further self-consistent field theory (\emph{e.g.}
  \cite{Nakamura2010}) or by determining fluctuations with respect to
  a reference system (\emph{e.g.}  \cite{Ermoshkin2004}) or through a
  mean-field treatment.  In principle, before our approximations at
  least, the field theory introduced here is not restricted to subsets
  of connectivities or specifically cyclized conformations nor is it
  \emph{a priori} a mean-field formalism.}

\bn{In the current calculation we have a single polymer chain that
  is}{ immersed in an ideal binary mixture of pointwise clustering
  centers with different functionality (number of links that the
  center can have with the polymer segments). As a mathematical device
  we can think of free segments of the chain being part of clusters of
  functionality one, {\sl i.e.} they cannot connect to any other
  segments.  Moreover the system is in solution with clustering
  centers of functionalities $a$ and $b\neq a$. The highly
  non-Gaussian field-theory resulting from the study of the model is
  quite complicated but can be approached through the saddle point
  approximation.  Assuming the polymer to be highly dense we can then
  use the Random Phase Approximation (RPA) to describe the polymer
  degrees of freedom. We are then able to extract the local densities
  of segments that form part of clusters of different sizes, the
  effective potential based on small density fluctuations around a
  background of a given density, and the static structure factor.}
{In the current treatment we develop the formalism and then
  investigate properties of the system in the scenario of the uniform
  polymer segment density with Gaussian fluctuations.  However it is
  also shown where this approximation breaks down.  One could
  certainly expect nonhomogenous phases to develop which can in
  principle also be addressed by the formalism together with the
  consideration of higher orders in density fluctuations
  \cite{Hong1981,Shi1996,Leibler1980}.}

{We shall refer to ``polymer segments'' as the \emph{monomeric units}
of which the polymer chain is built. ``Clustering centers'' refer 
to point-like seeds of ``clusters'' of segments of the polymer chain.
In other words the clustering centers function so as to attach to a
specific number of the segments of the polymer and in so doing to
localise these segments at a common point, binding them reversibly
into a cluster. Consequently a polymer segment is also the basic unit
to which a single attachment to a single cluster is possible.  
We shall deal with clustering centers of different functionalities
which form a multi-component system with the polymer and provide a
``sea'' of centers to form clusters with the polymer segments.}

{
Although our field-theoretical formulation includes no precise model
for the mechanism that causes clustering centers of a given
functionality to occur, we investigate in Section \ref{sec:janus} the
case where the functionalities (10 and 40) of the clustering centers
are the same as those determined for Janus particles in recent studies
\cite{Fantoni2011}.  Indeed there
has recently been much development in the techniques for the synthesis
of new patchy colloidal particles
\cite{Manoharan2003,Pawar2010,Glotzer2007,Zhang2004}.  One
particularly simple class of these anisotropic particles, called Janus
particles \cite{deGennes1992,Casagrande1989,Hong2006,Walther2008}, 
seem to form mainly clusters of either 10 or 40 particles.  Here Monte
Carlo simulations \cite{Sciortino2009,Giacometti2009} indicate that
mainly stable micelle (10 
particles) or vesicle (40 particles) arrangements of these particles
are to be found in the vapor phase.  Moreover it was found that the
clusters behave very similarly to an ideal gas, since 
the particles forming the cluster tend to arrange with their active
surfaces towards the cluster center.   

Janus chains have been suggested as potentially useful candidates 
for understanding interesting polymer phenomena
\cite{Ding2007}. We will apply our formalism to the case of a
dense polymer in a Janus fluid and in so doing we hope to
add to the recent interest for Janus particles interacting with
polymer chains \cite{Kim2009,Walther2008,Ding2007}. To the best of
our knowledge there are no results in the literature that prove the
clustering in the Janus fluid in the presence of the polymer. So we
will take as a working hypothesis the existence of such a
clustering, and make the approximation of treating the Janus fluid as
an ideal mixture of Janus partilces, micelles of Janus particles, and
vesicles of Janus particles (in the spirit of Ref. \cite{Fantoni2011}).
}

The paper is organized as follows: in Section \ref{sec:model} we
describe the model we are studying and formulate the field-theory, in
Section \ref{sec:spa} we perform the saddle point approximation, and
discuss when we expect the approximation to be most accurate. Section
\ref{sec:RPA} is devoted to an investigation of dense polymer system
with clustering.  We use the Random Phase Approximation and derive an
expression for the free energy density of the system and the effective
interaction caused by the clustering centers. In Section
\ref{sec:g(k)} we determine the structure factor and its curvature at
small wave vectors, in Section \ref{sec:janus} we finally solve the
Janus case numerically, and section \ref{sec:conclusions} is devoted
to the final remarks.   

\section{The model}
\label{sec:model}

The grand partition function for the ideal binary mixture of
clustering centers of functionality $a$ and of functionality $b$, can
be written as  
\bq \nonumber 
\Theta=\sum_{N_1=0}^\infty\sum_{N_a=0}^\infty\sum_{N_b=0}^\infty
\frac{(z_1 V)^{N_1}}{N_1!}
\frac{(z_a V)^{N_a}}{N_a!}
\frac{(z_b V)^{N_b}}{N_b!}
\eq
where we also allowed for a third species of particles, of
functionality one, which cannot cause aggregation. $z_i$ are the usual
fugacities for species $i$, V is the volume of
the system, and $N_1+N_a+N_b$ is the total number of clustering centers. 

We now wish to connect these to the polymer degrees of freedom \an{and
develop a partition function for the system of polymer together with
the clustering centers.}
{The ``clusters or clustering centers'' represents the free
{\sl macro-particles} making up the {\sl ideal mixture} (a two
component mixture with clusters of two different functionalities that
are living in a ``sea'' of clusters of functionality one: the
{\sl particles}) in which the 
polymer is immersed. These macro-particles are made up of a fixed
number of pointwise {\sl particles} ({\sl i.e.} they have fixed
functionality) each of which is {\sl linked} with one polymer
segment.} 

A suitable field-theoretic formalism was developed by Edwards
\cite{Edwards1988} to describe polymer gels.
Consider a field $\phi_1:\mathbb{R}^3\to\mathbb{R}$ then the following
Wick's theorem holds \an{(see Appendix \ref{app:GD})}
\bq \nonumber
I(\rr_1,\rr_2,\ldots,\rr_{2M})&=&{\cal N}\int[d\phi_1]\,
\phi_1(\rr_1)\phi_1(\rr_2)\ldots\phi_1(\rr_{2M})\,
e^{-\frac{1}{2}\int d\rr\,\phi_1^2(\rr)}\\
&=&\sum_{\mbox{all pairing}}
\delta(\rr_{l_1}-\rr_{l_2})\delta(\rr_{l_3}-\rr_{l_4})\cdots
\delta(\rr_{l_{2M-1}}-\rr_{l_{2M}})~,
\eq
where $l_i=1,2,\ldots,2M$ and $l_i\neq l_j$ for all $i\neq j$, \an{and
${\cal N}$ is the Gaussian normalization}.  

\begin{figure}[h!]
\begin{center}
\includegraphics[width=12cm]{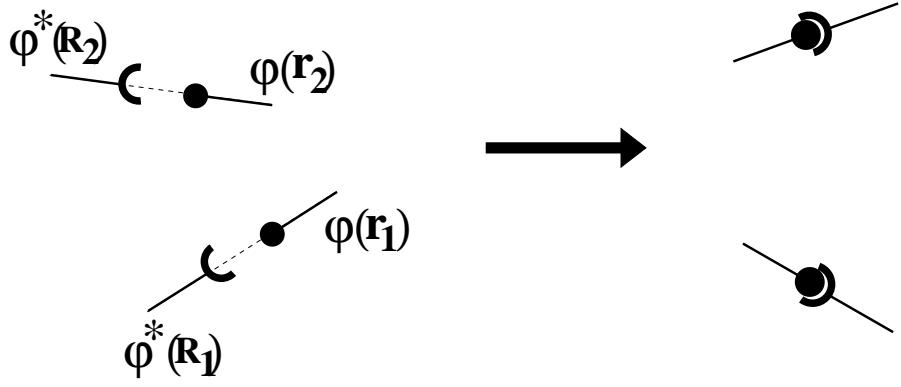}
\end{center}
\caption{A schematic representation of the role of the
field-theory. The fields $\varphi$ and $\varphi^\star$ are depicted
as functions of spatial variables. Multiplication by
$\exp(-\int\varphi\varphi^\star)$ and subsequent functional
integration enforces the linking of the spatial coordinates 
between pairs of $\varphi$ and $\varphi^\star$
(in all
possible ways).}
\label{fig:fields}
\end{figure}

If we introduce another field $\phi_2$, in terms of complex fields 
$\varphi=\phi_1+i\phi_2$ and $\varphi^\star=\phi_1-i\phi_2$, the
following identity follows (see Fig. \ref{fig:fields}),  
\bq \nonumber
J&=&{\cal N}^{'}\int [d\phi_1][d\phi_2]\,
\prod_{i=1}^M[\phi_1(\rr_i)+i\phi_2(\rr_i)]
\prod_{j=1}^{M^{'}}[\phi_1(\RR_j)-i\phi_2(\RR_j)]\,
e^{-\int d\rr\,\phi_1^2(\rr)-\int d\rr\,\phi_2^2(\rr)}\\\nonumber
&=&{\cal N}^{'}\int [d\varphi][d\varphi^\star]\,
\prod_{i=1}^M\varphi(\rr_i)
\prod_{j=1}^{M^{'}}\varphi^\star(\RR_j)\,
e^{-\int d\rr\,\varphi(\rr)\varphi^\star(\rr)}\\ \label{J}
&=&\delta_{M,M^{'}}\sum_{\mbox{all pairing}}
\delta(\rr_{l_1}-\RR_{m_1})\delta(\rr_{l_2}-\RR_{m_2})
\ldots\delta(\rr_{l_M}-\RR_{m_M})~,
\eq
where $l_i$ and $m_i$ can vary over $(1,2,\ldots,M)$ with $l_i\neq l_j$
and $m_i\neq m_j$ for all $i\neq j$.  This means that each $\varphi$
is associated with another $\varphi^\star$ through a Dirac delta
function in all possible pairwise combinations.  As shown in
Fig.~\ref{fig:fields}, we can view the fields $\varphi$ and
$\varphi^\star$ as being complementary, since the delta-function
connection does not occur between pairs of $\varphi$ or pairs of
$\varphi^\star$. This Gaussian theory, therefore, enumerates all
possible pairs of points $\rr_i$ and $\RR_j$ and enforces this by
inserting a Dirac delta.

\begin{figure}[h!]
\begin{center}
\includegraphics[width=10cm]{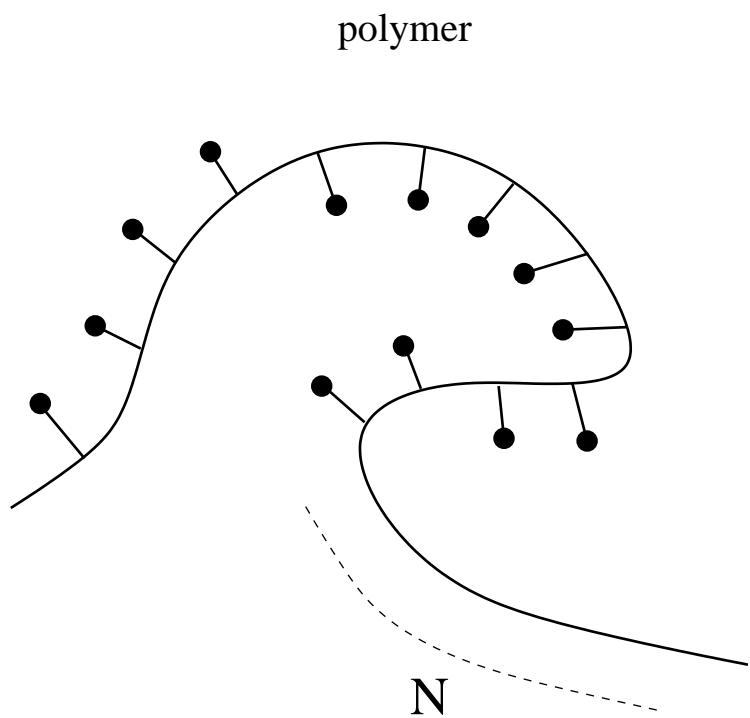}
\end{center}
\caption{Shows the polymer made up of $N$ equispaced links that are
susceptible to being linked into clusters.}
\label{fig:polymer}
\end{figure}

\begin{figure}[h!]
\begin{center}
\includegraphics[width=12cm]{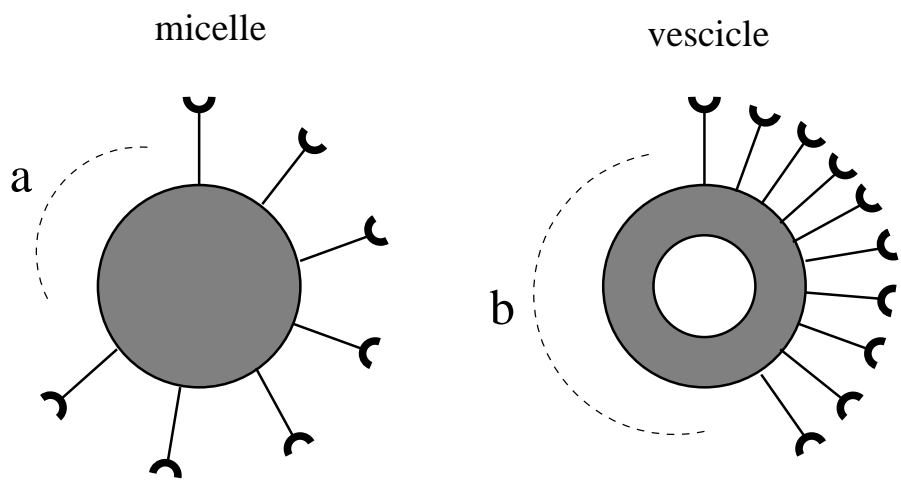}
\end{center}
\caption{Shows the clusters of Janus particles: the micelles are made of
$a=10$ links whereas the vesicles of $b=40$ links.}
\label{fig:clusters}
\end{figure}

We consider a polymer chain consisting of $N$ links (see
Fig. \ref{fig:polymer}). Given the Green 
function $G(\rr,\rr^{'})$ for the segment of chain between two
links, the conformation statistical weight of the polymer, \an{whose
conformation is described by the points $\{\RR_i\}$, is}
\bq
P(\{\RR_i\})=G(\RR_1,\RR_2)G(\RR_2,\RR_3)\ldots G(\RR_{N-1},\RR_N)~.
\eq
The polymer is immersed in an ideal mixture made up of two types
of cross-linked particles (see Fig. \ref{fig:clusters}) with
functionalities $a$ and $b$, respectively. We can also think of the
chain segments as forming two types of clusters (closed-shell
clusters, as for Janus particles \cite{Sciortino2009}) called micelles
and vesicles. What is important is the fixed functionality. The role
of these clustering centers is to provide 
the links of a certain fixed functionality between polymer segments,
thereby connecting a given number of segments of the polymer chain
together. Then the partition function can be written as
\bq \label{ZN}
Z_N=\sum_{N_1,N_a,N_b=0}^\infty {\cal Z}_{N_1,N_a,N_b}~,
\eq
where {$N=N_1+aN_a+bN_b$ is the total number of polymer segments or
of particles (since the field-theory requires that each segment must
be paired with a particle) and}
\bq\nonumber
{\cal Z}_{N_1,N_a,N_b}&=&{\cal N}\int d\RR_1\ldots d\RR_N
\an{e^{-v\sum_{n,m=1}^N\delta(\RR_n-\RR_m)}}
\int[d\varphi][d\varphi^\star]\,
e^{-\int d\rr\,\varphi(\rr)\varphi^\star(\rr)}\times\\ \nonumber
&&\varphi^\star(\RR_1)G(\RR_1,\RR_2)\varphi^\star(\RR_2)G(\RR_2,\RR_3)\ldots
G(\RR_{N-1},\RR_N)\varphi^\star(\RR_N)\times\\ \label{fieldtheory0}
&&\frac{1}{N_1!}\left(\int d\rr\,z_1\varphi(\rr)\right)^{N_1}
\frac{1}{N_a!}\left(\int d\rr\,z_a\varphi^a(\rr)\right)^{N_a}
\frac{1}{N_b!}\left(\int d\rr\,z_b\varphi^b(\rr)\right)^{N_b}~,
\eq
and the $z_i$ are generalized fugacities (that might also contain a
multiplicity associated with the functionality). {We have
explicitly added clusters of functionality one (the single particles)
here to represent the 
``sea'' of particles in which the polymer and the $a$ and $b$ clusters
are immersed. Clearly clusters of size one simply link to a
single polymer segment and therefore do not cause association as
the $a$ and $b$ clusters do. Note that we are not interested in
describing the precise model for the mechanism by which the clusters
of a given functionality  
are formed from the aggregation of particles (more on this in
Section \ref{sec:janus}); we just assume that this aggregation process
takes place}. At this level of description the physics 
of the precise mechanism for the multimerization resides in the
fugacities. \an{In our present formalism the clustering centers and
the particles are
pointlike.} In principle it is also possible to extend the current 
formalism to model clusters of finite extension. \an{We have also
added an excluded volume term $v$, to the polymer chain, with the
dimensions of a volume.} 

{It is possible to use the formalism without necessarily introducing 
the essentially inert clusters of functionality one, which turns out
to be the fugacity $z_1=1$ case of the equations derived below.  
Appendix~\ref{app:alternativefieldtheory} shows the details.
We continue with the slightly more general formalism here, noting
that $z_1\rightarrow 1$ will show no effects due to the addition
of these convenient clustering centers.}

We then find in a short hand notation, \an{and neglecting for the time
being the excluded volume term}, 
\bq \nonumber
Z_N&=&{\cal N}\int[d\varphi][d\varphi^\star]\left\{\prod d\RR\right\}
\left\{\prod G\right\}\eta^N 
\exp\left(-\int d\rr\, \varphi(\rr)\varphi^\star(\rr)+\right.\\
&&\left.\int d\rr\,\rho(\rr)\ln(\varphi^\star(\rr)/\eta)+
z_1\int d\rr\,\varphi(\rr)+z_a\int d\rr\,\varphi^a(\rr)+
z_b\int d\rr\,\varphi^b(\rr)\right)
\label{eq:fieldtheory}
\\
&=&{\cal N}\int[d\varphi][d\varphi^\star]\left\{\prod d\RR\right\}
\left\{\prod G\right\}\exp\{{\cal F}[\varphi,\varphi^\star]\}~,
\eq
where we introduced the microscopic density of polymer links
$\rho(\rr)=\sum_{i=1}^N\delta(\rr-\RR_i)$ and $\eta$ is an arbitrary
constant with the dimensions of a length to the power $-3/2$. In
the rest of the paper we will measure lengths in units of
$\eta^{-2/3}$. {A natural choice would be 
$\eta=\ell^{-3/2}$, with $\ell$ the Kuhn length of the polymer
segment.} 

\an{
\subsection{A simple example}

To clarify our formalism we here consider the simple example of a
polymer, with 4 polymer segments, interacting with two clustering
centers of functionality 
$a=2$. Using the properties of the Gaussian chains, the partition
function, neglecting the excluded volume term, is written as  
\bq \nonumber
{\cal Z}&=&{\cal N}\int[d\varphi][d\varphi^\star]\left\{\prod_{i=1}^4
d\RR_i\right\}\left\{\prod_{i=1}^{2}d\rr_i\right\}
\left\{\prod_{i=1}^{3}G(\RR_i,\RR_{i+1})\right\}
e^{-\int d\rr\,\varphi(\rr)\varphi^\star(\rr)}\times\\ \nonumber
&&\left\{\prod_{i=1}^4\varphi^\star(\RR_i)\right\}
z_2^2\varphi^2(\rr_1)\varphi^2(\rr_2)\\
&=&{\cal N}z_2^2\int\,d\rr_1d\rr_2\sum^\prime 
G(\rr_{i_1},\rr_{i_2})G(\rr_{i_2},\rr_{i_3})G(\rr_{i_3},\rr_{i_4})~.
\eq
where we used Eq. (\ref{J}) and the prime on the last summation symbol
indicates that we have to sum over all possible ways of assigning to
the indexes $(i_1,i_2,i_3,i_4)$ the set $(1,1,2,2)$. We then see that
the result is given by
\bq \nonumber
{\cal Z}&=&{\cal N}z_2^2\left[4\int\,d\rr_1d\rr_2
G(\rr_1,\rr_1)G(\rr_1,\rr_2)G(\rr_2,\rr_2)+
4\int\,d\rr_1d\rr_2
G(\rr_1,\rr_2)G(\rr_2,\rr_2)G(\rr_2,\rr_1)+\right.\\ \nonumber
&&\left. 4\int\,d\rr_1d\rr_2
G(\rr_1,\rr_2)G(\rr_2,\rr_1)G(\rr_1,\rr_2)\right]\\ \nonumber
&=&4\raisebox{-.6cm}{\includegraphics[width=6cm]{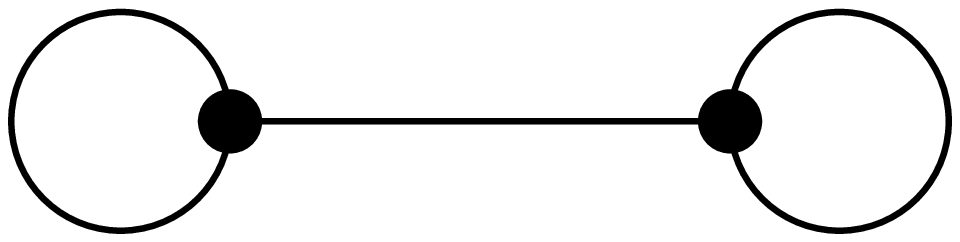}}+
4\raisebox{-1.5cm}{\includegraphics[width=5cm]{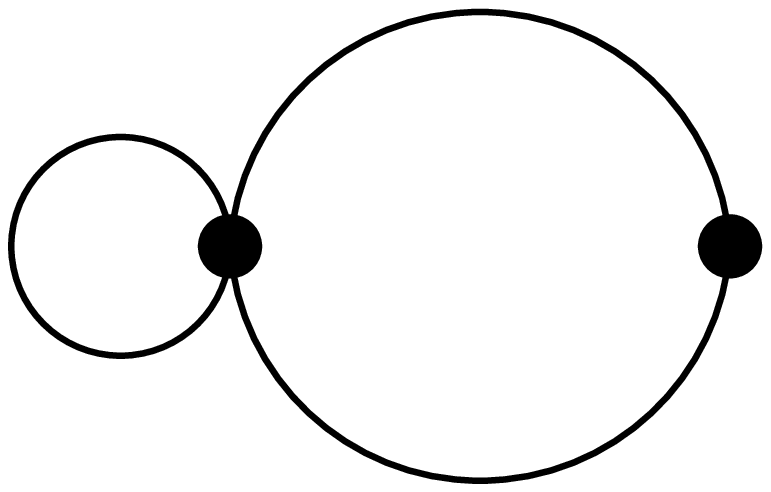}}+\\
&&4\raisebox{-1.6cm}{\includegraphics[width=4cm]{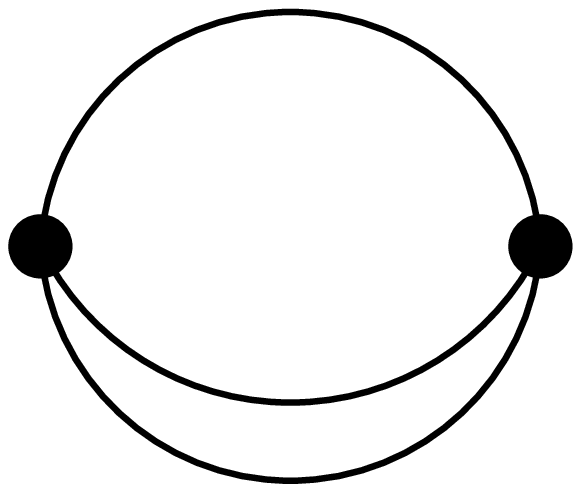}}~,
\eq
where lines represent polymer segments and black circles represent
integration. 
}
\section{The saddle point approximation} 
\label{sec:spa}
We look at the saddle point equations for the fields $\varphi$
and $\varphi^\star$ for any arbitrary yet fixed conformation of the
polymer $\rho=\rho(\rr) \ge 0$ $\forall \rr\in\mathbb{R}^3$.   
The saddle point approximation becomes necessary because the
field-theoretical formulation of the  
system, Eq.~(\ref{eq:fieldtheory}), is certainly highly non-Gaussian.
We proceed to calculate the saddle point and note that, for fixed
polymer conformation, the resulting condition is a set of algebraic
equations, {\sl i.e.} the saddle point depends only on the local
density. We shall also see that there exists always exactly one saddle
point, corresponding to the physical solution of the equations. 

We need to determine the critical point $(\bar{\varphi},\bar{\varphi}^\star)$, 
which amounts to solving 
\bq
\left.\frac{\delta{\cal F}}{\delta\varphi(\rr)}
\right|_{\bar{\varphi},\bar{\varphi}^\star}=0~,~~~
\left.\frac{\delta{\cal F}}{\delta\varphi^\star(\rr)}
\right|_{\bar{\varphi},\bar{\varphi}^\star}=0
\eq 
or 
\bq \label{sp1}
-\bar{\varphi}^\star+z_1+az_a\bar{\varphi}^{a-1}+bz_b\bar{\varphi}^{b-1}&=&0~,
\\ \label{sp2}
-\bar{\varphi}+\rho/\bar{\varphi}^\star&=&0~.
\eq
These can be combined to give
\bq \label{critical point}
-\rho/\bar{\varphi}+z_1+az_a\bar{\varphi}^{a-1}+bz_b\bar{\varphi}^{b-1}=0~.
\eq

\subsection{Properties of the solutions of the saddle point approximation}
\label{sec:propsol}
From Eq. (\ref{critical point}) we can write
\bq
-\rho+f(\bar{\varphi})=0~,
\eq
where $f(x)=z_1x+az_ax^a+bz_bx^b$. We can then show that there exists
at least one positive solution of Eq. (\ref{critical point}) for
$\rho>0$. Indeed when $\bar{\varphi}\to 0$ then
$-\rho+f(\bar{\varphi})=-\rho<0$, whereas when
$\bar{\varphi}>\rho/z_1$ then $-\rho+f(\bar{\varphi})>0$. Consequently
$-\rho+f(\bar{\varphi})$ goes through a zero when
$\bar{\varphi}\in[0,\infty)$. And it is easy to show that there is
exactly one positive solution $\bar{\varphi}>0$ for any $\rho>0$.
$f^\prime(\bar{\varphi})=z_1+a^2z_a\bar{\varphi}^{a-1}+b^2z_b\bar{\varphi}^{b-1}$
so $f^\prime(\bar{\varphi})>0$ for all $\bar{\varphi}\ge 0$. Since the
derivative is strictly positive and the polynomial goes from negative
to positive, there exists exactly one positive root $\bar{\varphi}$
for any $\rho>0$. As we show in Section \ref{sec:concentrations} the
local cluster densities of the saddle point are proportional to
$z_1\bar{\varphi}, az_a\bar{\varphi}^a,$ and
$bz_b\bar{\varphi}^b$, such that $\bar{\varphi}$ must be
non-negative.  

\subsection{Solutions for $\bar{\varphi}$ and $\bar{\varphi}^\star$}
Consider the particular case $a=2,b=0$, corresponding to reversible
crosslinking, then from Eq. (\ref{critical point}) we find for the
critical point 
\bq
\bar{\varphi}_\pm&=&\frac{-z_1\pm\sqrt{z_1^2+8z_2\rho}}{4z_2}~,\\
\bar{\varphi}^\star_\pm&=&\frac{4z_2\rho}{-z_1\pm\sqrt{z_1^2+8z_2\rho}}~,
\eq
which rules out the $\bar{\varphi}_-$ solution, as the logarithm of
$\bar{\varphi}_-^\star$ is not well defined.

\subsection{Expansion of ${\cal F}$ around the critical point}
We can expand the function ${\cal F}$ around the critical point to
second order in the fluctuations
$\Delta\varphi=\varphi-\bar{\varphi}=\Delta\phi_1+i\Delta\phi_2$ and 
$\Delta\varphi^\star=\varphi^\star-\bar{\varphi}^\star=\Delta\phi_1-i\Delta\phi_2$, 
\bq \nonumber
{\cal F}[\bar{\varphi}+\Delta\varphi,\bar{\varphi}^\star+\Delta\varphi^\star]
&=&{\cal F}[\bar{\varphi},\bar{\varphi}^\star]+\\ \nonumber
&&\frac{1}{2}\int d\rr_1\int d\rr_2
\left(\begin{array}{cc}
\Delta\phi_1(\rr_1) &\Delta\phi_2(\rr_1)
\end{array}\right)\cdot
F_2(\rr_1,\rr_2)\cdot
\left(\begin{array}{c}
\Delta\phi_1(\rr_2)\\
\Delta\phi_2(\rr_2)
\end{array}\right)+\\
&&\mbox{third order terms}~,
\eq
where $F_2(\rr_1,\rr_2)$ is the following $2\times 2$ matrix
\bq
F_2(\rr_1,\rr_2)=
\left(\begin{array}{cc}
\left.\frac{\delta^2{\cal F}}{\delta\phi_1(\rr_1)\delta\phi_1(\rr_2)}
\right|_{\bar{\varphi},\bar{\varphi}^\star} &
\left.\frac{\delta^2{\cal F}}{\delta\phi_1(\rr_1)\delta\phi_2(\rr_2)}
\right|_{\bar{\varphi},\bar{\varphi}^\star} \\
\left.\frac{\delta^2{\cal F}}{\delta\phi_2(\rr_1)\delta\phi_1(\rr_2)}
\right|_{\bar{\varphi},\bar{\varphi}^\star} &
\left.\frac{\delta^2{\cal F}}{\delta\phi_2(\rr_1)\delta\phi_2(\rr_2)}
\right|_{\bar{\varphi},\bar{\varphi}^\star}
\end{array}\right)~.
\eq
 
We have
\bq
\left.\frac{\delta^2{\cal F}}{\delta\phi_1(\rr_1)\delta\phi_1(\rr_2)}
\right|_{\bar{\varphi},\bar{\varphi}^\star}&=&
\left[-2-\frac{\rho}{\bar{\varphi}^{\star 2}}+a(a-1)z_a
\bar{\varphi}^{a-2}+b(b-1)z_b\bar{\varphi}^{b-2}\right]\delta(\rr_1-\rr_2)~,\\
\left.\frac{\delta^2{\cal F}}{\delta\phi_1(\rr_1)\delta\phi_2(\rr_2)}
\right|_{\bar{\varphi},\bar{\varphi}^\star}&=&
i\left[\frac{\rho}{\bar{\varphi}^{\star 2}}+a(a-1)z_a
\bar{\varphi}^{a-2}+b(b-1)z_b\bar{\varphi}^{b-2}\right]\delta(\rr_1-\rr_2)~,\\
\left.\frac{\delta^2{\cal F}}{\delta\phi_2(\rr_1)\delta\phi_2(\rr_2)}
\right|_{\bar{\varphi},\bar{\varphi}^\star}&=&
\left[-2+\frac{\rho}{\bar{\varphi}^{\star 2}}-a(a-1)z_a
\bar{\varphi}^{a-2}-b(b-1)z_b\bar{\varphi}^{b-2}\right]\delta(\rr_1-\rr_2)~,
\eq
We can then introduce the matrix ${\cal M}[\rho]$ as
\bq \mbox{\footnotesize
$\left(
\begin{array}{cc}
-2-\frac{\rho}{\bar{\varphi}^{\star 2}}+a(a-1)z_a
\bar{\varphi}^{a-2}+b(b-1)z_b\bar{\varphi}^{b-2} & 
i\left[\frac{\rho}{\bar{\varphi}^{\star 2}}+a(a-1)z_a
\bar{\varphi}^{a-2}+b(b-1)z_b\bar{\varphi}^{b-2}\right]\\
i\left[\frac{\rho}{\bar{\varphi}^{\star 2}}+a(a-1)z_a
\bar{\varphi}^{a-2}+b(b-1)z_b\bar{\varphi}^{b-2}\right]& 
-2+\frac{\rho}{\bar{\varphi}^{\star 2}}-a(a-1)z_a
\bar{\varphi}^{a-2}-b(b-1)z_b\bar{\varphi}^{b-2}
\end{array}
\right)~,$}
\eq
Then the partition function can be rewritten as 
\bq \label{sppf}
Z_N\approx{\cal N}^{'}\int\left\{\prod d\RR\right\}\left\{\prod G\right\}
\exp\{M[\rho]\}~,
\eq
where 
\bq \label{M}
M[\rho]={\cal F}[\bar{\varphi},\bar{\varphi}^\star]-
\frac{1}{2}\int d\rr\,\ln\left\{\det {\cal M}[\rho]\right\}~,
\eq
and
\bq
\det {\cal M}[\rho]&=&4\left[1+z_aa(a-1)
\bar{\varphi}^{a}/\rho+z_bb(b-1)\bar{\varphi}^{b}/\rho\right]~.
\eq
For finite $\rho$ the determinant is always positive with positive
real parts of the eigenvalues, if $\bar{\varphi} > 0$. Hence the saddle
point is stable.

As the density $\rho$ increases we note that the free energy
calculation due to the saddle point ${\cal
  F}[\bar{\varphi},\bar{\varphi}^\star]$ grows at least as $\rho$, but
that the fluctuation contribution $-
\frac{1}{2}\int d\rr\,\ln\left\{\det {\cal M}[\rho]\right\}$ strives
to a constant. Therefore at sufficiently high density we expect the
relative contribution of the fluctuations to become negligible.

In the particular case $a=2,b=0$ we find
\bq
\det {\cal M}_{\pm}[\rho]&=&4\left(1+2z_2\bar{\varphi}_{\pm}/
\bar{\varphi}_{\pm}^\star\right)~.
\eq

\section{The Random Phase Approximation}
\label{sec:RPA}

In this section we express the polymer in terms of a segment
density via the so-called random phase approximation.  We
restrict ourselves to the situations where the
polymer segments are presumed to be distributed homogeneously 
with Gaussian density fluctuations around this value. 
For sufficiently dense systems this has been treated as a reasonable
approximation \cite{Deam1976} in permanently networked systems.
We do not investigate the cases of possible inhomogeneous phases.  
In principle we would then need to expand our results in the preceding
sections to higher orders in the density fluctuations
\cite{Hong1981,Shi1996,Leibler1980} or attempt  
to express our results in terms of more complex quantities 
\cite{Haronska1994a,Vilgis1994a}.  However, we do
investgate where our RPA with the homogeneity assumption fails as one
indicator for possibly different phase behaviour in the system.

\subsection{Basic formulation}
Clearly our clustering formalism produces a significantly nontrivial
density dependence. Presuming that our system behaves like a highly
dense polymer melt, where the fluctuations of
$\rho=\bar{\rho}+\Delta\rho$ are small, one can use the following
Random Phase Approximation \cite{Leibler1980} (RPA) (see Appendix
\ref{app:RPA}), 
\bq \nonumber
\int\left\{\prod d\RR\right\}\left\{\prod G\right\}\ldots&=&
{\cal N^{''}}\int[d\Delta\rho]\exp\left(-\frac{1}{2}
\int d\rr\int d\rr^{'}\,\Delta\rho(\rr)
\frac{\widehat{\tilde{S}_0^{-1}}(|\rr-\rr^{'}|)}{V}
\Delta\rho(\rr^{'})\right)\ldots\\ \label{eq:rpa}
&=&{\cal N^{''}}\int[d\Delta\tilde{\rho}]\exp\left(
-\frac{1}{2}\frac{1}{V}\sum_{\kk}\Delta\tilde{\rho}(\kk)
\frac{\tilde{S}_0^{-1}(k)}{V}
\Delta\tilde{\rho}(-\kk)\right)\ldots~,
\eq 
where we denoted with a tilde the Fourier transform \cite{FT} and with
a hat an inverse Fourier transform. We note that this type of approach
is not atypical in calculations for quenched gels \cite{Nedebock1999}.

Expanding to second order in the density fluctuations we find, from
Eq. (\ref{M}) and Eqs. (\ref{sp1})-(\ref{sp2}),

\bq \label{ABC}
M[\bar{\rho}+\Delta\rho]&=&
AV+B\int d\rr\,\Delta\rho(\rr)+C\int d\rr\,[\Delta\rho(\rr)]^2+\ldots
\\ \nonumber
&=&AV+C\int d\rr\,[\Delta\rho(\rr)]^2+\ldots~,
\eq
where we used the fact that $\int d\rr\,\Delta\rho(\rr)=0$. $V$
is the volume of the box, $A,B,$ and $C$ are given functions 
of $z_1,z_2,$ and $\bar{\rho}$ and for the particular case $a=2,b=0$
can be found in Appendix \ref{app:AB1}. Notice that in this case $A_-$
is not defined for any values of the average density so only the
$A_+,B_+,$ and $C_+$ solution is physically meaningful, {\sl i.e.}
they correspond to the expected positive $\bar{\varphi}$ solution (see
Section \ref{sec:propsol}). 

We then obtain the following approximation for the partition function
of Eq. (\ref{sppf})
\bq \label{RPA}
Z_N\approx{\cal N^{'''}}e^{AV-
\frac{1}{2}\int d\rr\,\ln(
\widehat{\tilde{\cal S}_0^{-1}}/V-2C)}~,
\eq
where $\widehat{\tilde{\cal S}_0^{-1}}$ is the operator whose
$\rr,\rr^\prime$ component is given by
$\widehat{\tilde{S}_0^{-1}}(|\rr-\rr^\prime|)$.

In terms of the free energy density $\beta f=-\ln(Z_N)/V$ we find, in
the thermodynamic limit ($V\to\infty$ with $\bar{\rho}=N/V$ constant),  
\bq
\beta f=-A-\frac{1}{2}\ln(-2C)~.
\eq

\subsection{Local clustered segment densities}
\label{sec:concentrations}
Following the usual method for grand-canonical ensemble it is possible
to calculate the local densities of segments that form part of
different sizes of clusters. We then compute
\bq
n_1=\frac{z_1}{N}\frac{\partial\ln Z_N}{\partial z_1}~,
\eq
for segments of the chain that are not cross-linked, and
\bq
n_x=x\frac{z_x}{N}\frac{\partial\ln Z_N}{\partial z_x}~,
\eq
for the density of segments part of clusters of size $x$.

Neglecting the logarithmic corrections
due to the quadratic fluctuations, we have
\bq
\ln Z_N=VA=M[\bar{\rho}]~.
\eq
Within the saddle point approximation we split the contributions into
parts due to the saddle point above and due to the quadratic
fluctuations (notice that 
this analysis holds also locally at the level of the partition
function for an arbitrary $\rho=\rho(\rr)$ profile of polymer chain
density) 
\bq \nonumber
n_x&=&x\frac{z_x}{N}\frac{\partial{\cal F}
[\bar{\varphi},\bar{\varphi}^\star]|_{\rho=\bar{\rho}}}{\partial z_x}
-x\frac{z_xV}{2N}\frac{\partial\ln\det{\cal M}[\bar{\rho}]}{\partial z_x}\\
&=&n^S_x+n^Q_x~,
\eq
where $x=1, x=a,$ or $x=b$. We then find
\bq
n^S_x=\frac{xz_x\bar{\varphi}^x}{\bar{\rho}}~.
\eq
Since $n^S_x$ have to be real and non-negative, in the saddle point
approximation, also the solutions $\bar{\varphi}$ have to be real and
non-negative. Immediately from the saddle point equation (\ref{critical
  point}) follows that
\bq \label{cl1}
n^S_1+n^S_a+n^S_b=1~,
\eq
must hold generally. We also find after some algebra
\bq \label{cl2}
n^Q_1+n^Q_a+n^Q_b=0~.
\eq
This means that the saddle point approximation conserves the total
number of segments for {\sl any} density. Consequently, any average
over these density dependent expressions, irrespective of the
approximation, must satisfy the conservation. However we note that the
conservation laws (\ref{cl1}) and (\ref{cl2}) do not prevent possibly
negative $n^S_x+n^Q_x$ which can arise in the region where the
fluctuation part is not sufficiently smaller than the saddle point. As
the validity of the saddle point improves with density, also this
possibility disappears. 

For the special case $a=2,b=0$ we have
\bq \label{n1s}
n^S_1&=&\frac{2z_1}{z_1+\sqrt{z_1^2+8z_2\bar{\rho}}}~,\\
n^Q_1&=&\frac{z_1(-z_1+\sqrt{z_1^2+8z_2\bar{\rho}})}
{2\bar{\rho}(z_1^2+8z_2\bar{\rho})}~,
\eq
as a consequence we see that the fraction of monomers not in a
cross-link decreases with the density.

\subsection{The effective potential}
\label{sec:effective}

Upon integrating over the degrees of freedom associated with the clustering 
centers the remaining integral in the partition function is that 
over the polymer density degrees of freedom (in the RPA).  This 
permits us to interpret the effective interaction between polymer
segments as caused by the clustering.
It consists of the typically attractive contribution to the polymer-polymer
quadratic density fluctuations from the aggregating fields 
and any direct polymer-polymer interaction (such as excluded volume 
interactions).

From Eq. (\ref{sppf}) and (\ref{eq:rpa})-(\ref{ABC}) we can rewrite
the partition function as 
\bq
Z_N\approx {\cal N}^{'''}\int[d\Delta\rho]
e^{-\frac{1}{2}\int d\rr\int d\rr^{'}\,\Delta\rho(\rr)
\frac{\widehat{\tilde{S}_0^{-1}}(|\rr-\rr^{'}|)}{V}
\Delta\rho(\rr^{'})}
e^{AV}e^{-\frac{1}{2}\int d\rr\int d\rr^{'}\,\Delta\rho(\rr)W(\rr-\rr^{'})
\Delta\rho(\rr^{'})}
\eq
where the effective potential {between the polymer segments}, $W$,
is given by 
\bq
W(\rr-\rr^\prime)=-
\left.\frac{\delta^2 M[\rho]}{\delta\rho(\rr)\delta\rho(\rr^\prime)}
\right|_{\rho(\rr)=\bar{\rho}}=
-2C(z_1,z_a,z_b;\bar{\rho})\delta(\rr-\rr^\prime)~.
\eq

We can then split the contribution from the saddle point and the
quadratic contribution of Eq. (\ref{M}) and write
\bq
C&=&C_S+C_Q~,\\
C_S&=&\frac{1}{2}\left.\frac{\partial^2
f^S(\bar{\varphi},\rho)}{\partial
  \rho^2}\right|_{\rho=\bar{\rho}}~,\\
C_Q&=&\frac{1}{2}\left.\frac{\partial^2
f^Q(\bar{\varphi},\rho)}{\partial
  \rho^2}\right|_{\rho=\bar{\rho}}~,
\eq
where
\bq
f^S&=&-\rho+\rho\ln(\rho/\bar{\varphi})+z_1\bar{\varphi}+z_a\bar{\varphi}^a+
z_b\bar{\varphi}^b~,\\
f^Q&=&-\frac{1}{2}\ln[4(1+a(a-1)z_a\bar{\varphi}^a/\rho+b(b-1)
\bar{\varphi}^b/\rho)]~.
\eq
Now, using Eq. (\ref{critical point}), we find $\partial f^S/\partial
\rho=\ln\rho-\ln\bar{\varphi}$ and ,using the property 
$\partial\bar{\varphi}/\partial\rho=1/(z_1+ a^2z_a\bar{\varphi}^{a-1}+
b^2z_b\bar{\varphi}^{b-1})$, follows $\partial^2 f^S/\partial
\rho^2=1/\rho-1/(z_1\bar{\varphi}+ z_aa^2\bar{\varphi}^a+
z_bb^2\bar{\varphi}^b)$. Let us assume for definiteness that
$b>a$. Then when $\rho$ is very small $z_1\bar{\varphi}\approx\rho$ and 
$\partial^2 f^S/\partial\rho^2\approx a(a-1)(z_a/z_1^a)
\rho^{a-2}/[1-a(z_a/z_1^a)\rho^{a-1}]$, while when $\rho$ is
very large $z_bb\bar{\varphi}^b\approx\rho$ so that $\partial^2 f^S/\partial
\rho^2\approx (b-1)/(b\rho)$. Moreover we find in the large $\rho$ limit
that $\partial^2 f^Q/\partial\rho^2$ behaves at least as
$1/\rho^2$. 

We remark that in the small $\rho$ limit in the $a=2,b=0$ case $\partial^2
f^Q/\partial\rho^2\approx 10z_a^2/z_1^4$ whereas in the $a=10,b=40$
case $\partial^2 f^Q/\partial\rho^2\approx
-3240z_a\rho^7/z_1^{10}$. However we guard against interpreting this
as a repulsive interaction as the saddle point approximation to our
field-theory is not expected to be accurate at small densities. This
repulsive contribution in the small 
density limit for the effective potential of the Janus case (see
Section \ref{sec:janus}) explains the fact that here the RPA can be
valid (see Section \ref{sec:validity}) even if we do not add any
excluded volume interaction to the polymer. 

For the simple case $a=2,b=0$ we then find that $C_+=C_S+C_Q$ where
\bq
C_S&=&\frac{1}{4\bar{\rho}}\left(1-\frac{z_1}
{\sqrt{z_1^2+8z_2\bar{\rho}}}\right)~,\\
C_Q&=&\frac{1}{8\bar{\rho^2}}
\left[\frac{64(\bar{\rho}z_2)^2}{(z_1^2+8z_2\bar{\rho})^2}-1+
\frac{z_1^3+12z_1z_2\bar{\rho}}{(z_1^2+8z_2\bar{\rho})^{3/2}}
\right]~.
\eq

Here we also find in the small $\bar{\rho}$ limit
\bq
C_S&=&\frac{z_2}{z_1^2}-\frac{6z_2^2}{z_1^4}\bar{\rho}+O(\bar{\rho}^2)~,\\
C_Q&=&\frac{5z_2^2}{z_1^4}-\frac{88z_2^3}{z_1^6}\bar{\rho}+O(\bar{\rho}^2)~,
\eq
which tells us that the energy to cross-link goes to a constant
proportional to $z_2$.

In the large $\bar{\rho}$ limit
\bq \label{C1l}
C_S&=&\frac{1}{4\bar{\rho}}-\frac{z_1}{8\sqrt{2z_2}\bar{\rho}^{3/2}}
+\frac{z_1^3}{128\sqrt{2z_2^3}\bar{\rho}^{5/2}}+O(\bar{\rho}^{-3})~,\\
C_Q&=&\frac{3z_1}{32\sqrt{2z_2}\bar{\rho}^{5/2}}+O(\bar{\rho}^{-3})~,
\eq
which tells us that the energy to cross-link goes to zero as
$1/\bar{\rho}$, in accord with the fact that we are in a dense system. 

The effective potential calculated here, based on small density
fluctuations around a background of a given density, is dependent on
the average density. As expected the clustering produces a local
attractive interaction. 

However, it is interesting to note that the strength of this interaction
decreases with average density. We attribute this to the fact that the
fraction of free segments ({\sl i.e}. those in clusters of size 1) decreases
with the average density according to Eq. (\ref{n1s}). Therefore, for
large $\bar{\rho}$, the number of additional free segments gained by
increasing the density from $\bar{\rho}$ to $\bar{\rho}+\Delta$ is
proportional to $\bar{\rho}^{-1/2}$ leading to a pairwise contribution
$\Delta^2/\bar{\rho}$ as found in Eq. (\ref{C1l}).

\subsection{Validity of RPA}
\label{sec:validity}
The RPA is based on a homogeneity assumption which no longer holds
when the RPA itself predicts overly large fluctuations.  
In order to obtain Eq. (\ref{RPA}) we must have that
$\tilde{S}_0^{-1}(k)/V-2C$ is a strictly positive function for all
values of the wave vector $k$. Since $\tilde{S}_0^{-1}(k)$ is a
monotonically increasing function of $k$, the RPA will be valid as long
as 
\bq
C<\frac{\tilde{S}_0^{-1}(0)}{2V}=\frac{1}{2V\bar{\rho}^2}~.
\eq
In the thermodynamic limit one would require that $C<0$, for the
validity of RPA. 

We can then extend the region of the validity of RPA by adding an
excluded volume effect \cite{DoiEdwards} to the polymer which amounts
to taking 
$M[\rho]\rightarrow M[\rho]-v\int d\rr\,\rho^2(\rr)$ with $v$ a
positive constant with the dimensions of a volume. We will then have 
\bq \label{excluded volume}
A\to A-\bar{\rho}^2v~,~~~B\to B-2\bar{\rho}v~,~~~C\to C-v~,
\eq
and the validity of RPA becomes $C<v$.

For the $a=2,b=0$ case we have that $C_+$ is always positive so the
RPA cannot be applied without the excluded volume interaction. As a
matter of fact we have $\lim_{\bar{\rho}\to\infty}C_+=0$ and
$\lim_{\bar{\rho}\to 0}C_+=z_2(z_1^2+5z_2)/z_1^4$ and $C_+$ is a
monotonically decreasing function of $\bar{\rho}$. So by choosing $v$
any arbitrarily small positive constant we are able to extend the
range of validity of RPA to arbitrarily large densities.

\begin{figure}[h!]
\begin{center}
\includegraphics[width=12cm]{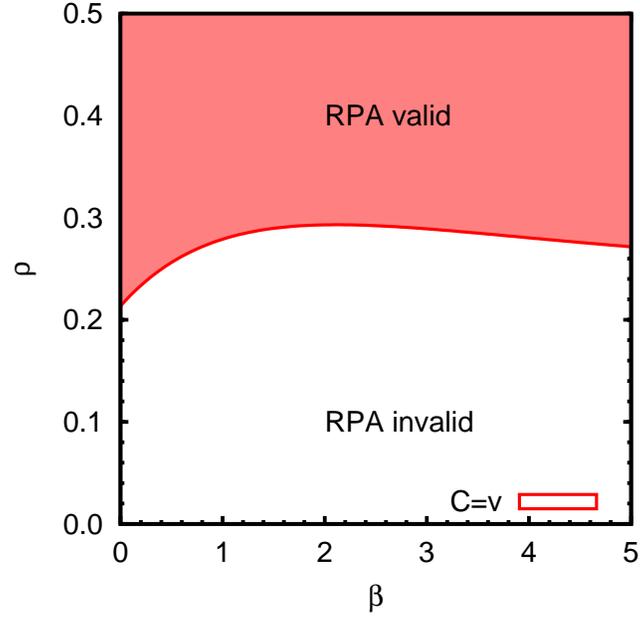}
\end{center}
\caption{(Color online) Shows the RPA validity region of the phase diagram, in the
$a=2,b=0$ case, for $z_1=1,z_2=e^{2\beta}$ and $v=1$.}
\label{fig:vals}
\end{figure}

In this case choosing $z_1=1,z_2=e^{2\beta}$, and $v=1$ the validity
domain in the phase diagram is determined in Fig. \ref{fig:vals}.

\begin{figure}[h!]
\begin{center}
\includegraphics[width=12cm]{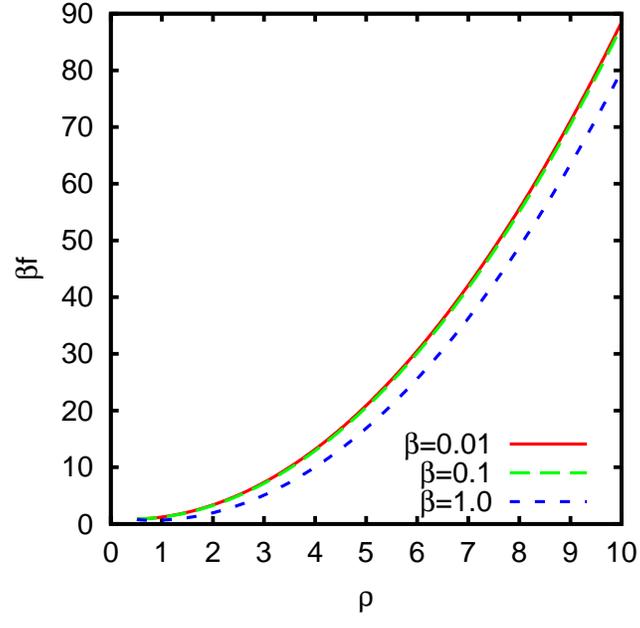}
\end{center}
\caption{(Color online) Shows the free energy density as a function of the average
density in the $a=2,b=0$ case, for $z_1=1,z_2=e^{2\beta}$ and $v=1$.} 
\label{fig:f}
\end{figure}

In Fig. \ref{fig:f} we show the behavior of the free energy density as
a function of density in the case $a=2,b=0$. Here we choose
$z_1=1,z_2=e^{2\beta}$ were $\beta=1/k_\text{B}T$ with $k_\text{B}$ Boltzmann
constant and $T$ is the temperature and $v=1$. 

\section{The static structure factor}
\label{sec:g(k)}

The Fourier transform of the pair correlation function is defined as
\cite{Doi} 
\bq
\tilde{g}(\kk)=
\frac{1}{N}\langle\tilde{\rho}(\kk)\tilde{\rho}(-\kk)\rangle=
\frac{1}{N}\langle\Delta\tilde{\rho}(\kk)\Delta\tilde{\rho}(-\kk)\rangle
~,~~~\kk\neq 0~.
\eq
The quantity $\tilde{g}(\kk)$ can be measured experimentally by light
scattering. Moreover one can extract some important information on the
polymer properties from the small $k=|\kk|$ behavior: 
\bq
\tilde{g}(\kk)=\tilde{g}(0)\left(1-\frac{k^2}{3}R_g^2+\ldots\right)~,
\eq
where $R_g$ is the radius of gyration of the polymer, namely
\bq
R_g^2=\frac{1}{2N^2}\sum_{n=1}^N\sum_{m=1}^N\langle(\RR_n-\RR_m)^2\rangle~.
\eq

Now using the result from the RPA we find
\bq
N\tilde{g}(k)=V[\tilde{S}_0^{-1}(k)/V-2C]^{-1}~,
\eq
which when $C=-v$, agrees with Edwards' result
\cite{DoiEdwards,Shimada1988} for polymer chains with only excluded
volume interactions \cite{pcsf}. Notice that the effective
potential $C$ is in general a function of $\bar{\rho},
z_1,z_a,$ and $z_b$. As shown in Section \ref{sec:effective}, $C+v$
tends to be positive (attractive interaction between polymer segments)
in the presence of clustering centers. So we expect there to be a
regime of density for which there is a balance between the repulsion
due to the excluded volume effect and the attraction due to clustering
making $C\approx 0$. In such case our result reproduces the one for
the ideal chain (see section 1.2.3 in Ref. \cite{Doi}).

In the small $k$ limit we find
\bq
N\tilde{g}(k)=\frac{V^2}{1/\bar{\rho}^2-2CV}
\left[1-\frac{k^2}{3}\xi^2+\ldots\right]~,
\eq
where $\bar{\rho}=N/V$ is the average polymer segment density for a
single long polymer chain and the ``curvature'' of the structure
factor at $k=0$ is   
\bq
\xi^2=\frac{\ell^2\bar{\rho} V}{6-12\bar{\rho}^2 CV}~,
\eq
where $\ell$ is the Kuhn length of the polymer. And in the
thermodynamic limit 
\bq
(\xi/\ell)^2\to -\frac{1}{12\bar{\rho}C}~.
\eq
So that at large polymer densities the curvature tends to a
constant (we note that $C$ now includes the excluded volume as in
Eq. (\ref{excluded volume})).  

We also find, in the thermodynamic limit, the following expression for
the structure factor
\bq
g(k)\to\frac{12}{(\ell k)^2-24\bar{\rho}C}~.
\eq
Notice that in the absence of the effective interaction ($C=0$) the
structure factor diverges at $k=0$.

In the $a=2,b=0$ case, at constant $V$, in the small $N$ limit we find
the free polymer result $\xi^2=N\ell^2/6+O(N^3)$. In the large $N$
limit we find  
\bq \label{limitrg}
(\xi/\ell)^2=\frac{1}{12v}\frac{V}{N}+\frac{1}{48v^2}\frac{V^2}{N^2}
+O(1/N^{5/2})~. 
\eq
Given the densest possible filling $N/V \sim 1/v$ the curvature 
tends to a constant.

\section{The Janus case}
\label{sec:janus}

Although our field-theoretical formulation includes no precise model
for the mechanism that causes clustering centers of a given
functionality to occur, we investigate here the case where the
functionalities (10 and 40) of the clustering centers are the same as
those determined for Janus particles in recent studies.  Indeed there
has recently been much development in the techniques for the synthesis
of new patchy colloidal particles
\cite{Manoharan2003,Pawar2010,Glotzer2007,Zhang2004}.  One
particularly simple class of these anisotropic particles, called Janus
particles \cite{deGennes1992,Casagrande1989,Hong2006,Walther2008}, 
seem to form mainly clusters of either 10 or 40 particles.  Here Monte
Carlo simulations \cite{Sciortino2009,Giacometti2009} indicate that
mainly stable micelle (10 
particles) or vesicle (40 particles) arrangements of these particles
are to be found in the vapor phase.  Moreover it was found that the
clusters behave very similarly to an ideal gas, since 
the particles forming the cluster tend to arrange with their active
surfaces towards the cluster center.   

Janus chains have been suggested as potentially useful candidates 
for understanding interesting polymer phenomena
\cite{Ding2007}. We will apply our formalism to the case of a
dense polymer in a Janus fluid and in so doing we hope to
add to the recent interest for Janus particles interacting with
polymer chains \cite{Kim2009,Walther2008,Ding2007}. To the best of
our knowledge there is no results in the literature that proves the
clustering in the Janus fluid in the presence of the polymer. So we
will take as a working hypothesis the existence of such a
clustering. And make the approximation of treating the backbone units
of the polymer (the Janus particles) as an ideal fluid.

Given the general setting described above we can apply our
theoretical model to a polymer in a Janus fluid. By this we think of
chain segments only clustering to form limited closed shell
conformations, {\sl i.e.} micelles and vesicles. As mentioned before,
we however do not consider the nature of spacial extent of the
clustering in detail and simply presume that it still occurs in the
same way as if the Janus particles where not connected to the polymer.

In the Janus case we have to choose $a=10$ and $b=40$ (see
Fig. \ref{fig:clusters}). We then find 
for the determination of the critical point an algebraic equation of
degree $40$, Eq. (\ref{critical point}). As expected, this equation has
just one solution for which $A$ (from Eq. (\ref{ABC})) is real and
non-negative. 

We can see the generalized fugacities defined as $z_i\propto\exp(-\beta
u_i+\beta\mu_i)$ for $i=1,10,40$, where $u_i$ is the average internal
energies of the cluster of $i$ Janus particles and $\mu_i$ is the chemical
potential of this cluster species. It is moreover reasonable to take
$\mu_i\approx\mu$ independent of $i$ ($\mu$ being the chemical
potential of the vapor phase of the Janus fluid) so that we get
$z_i\propto\exp(-\beta u_i)$.
 
\begin{figure}[h!]
\begin{center}
\includegraphics[width=12cm]{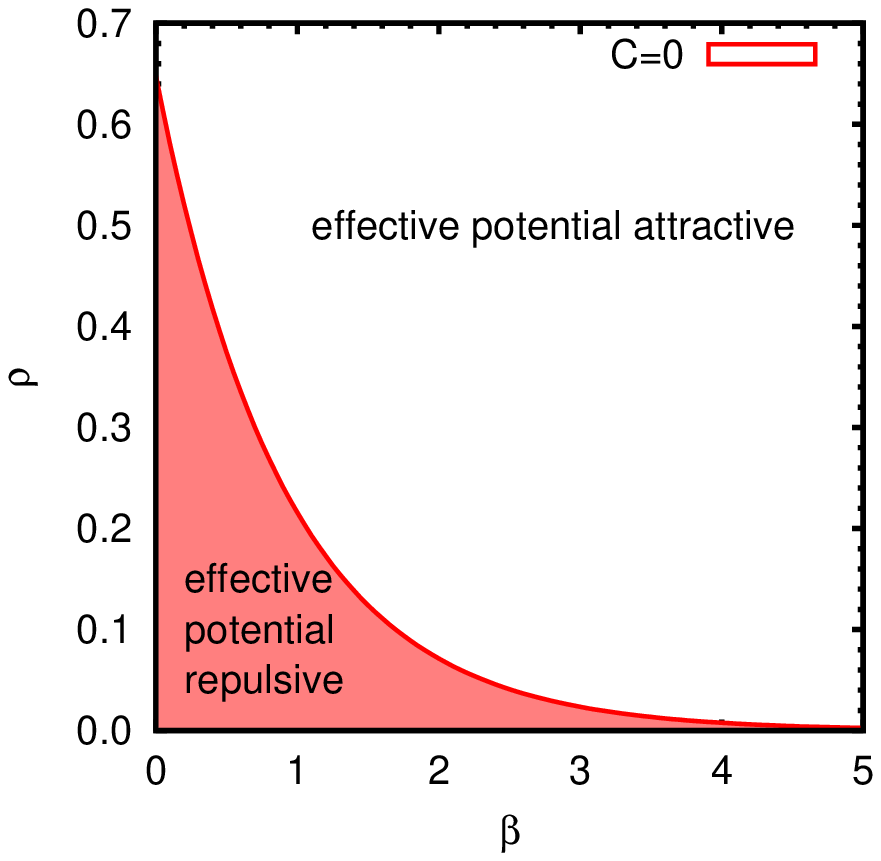}\\
\includegraphics[width=12cm]{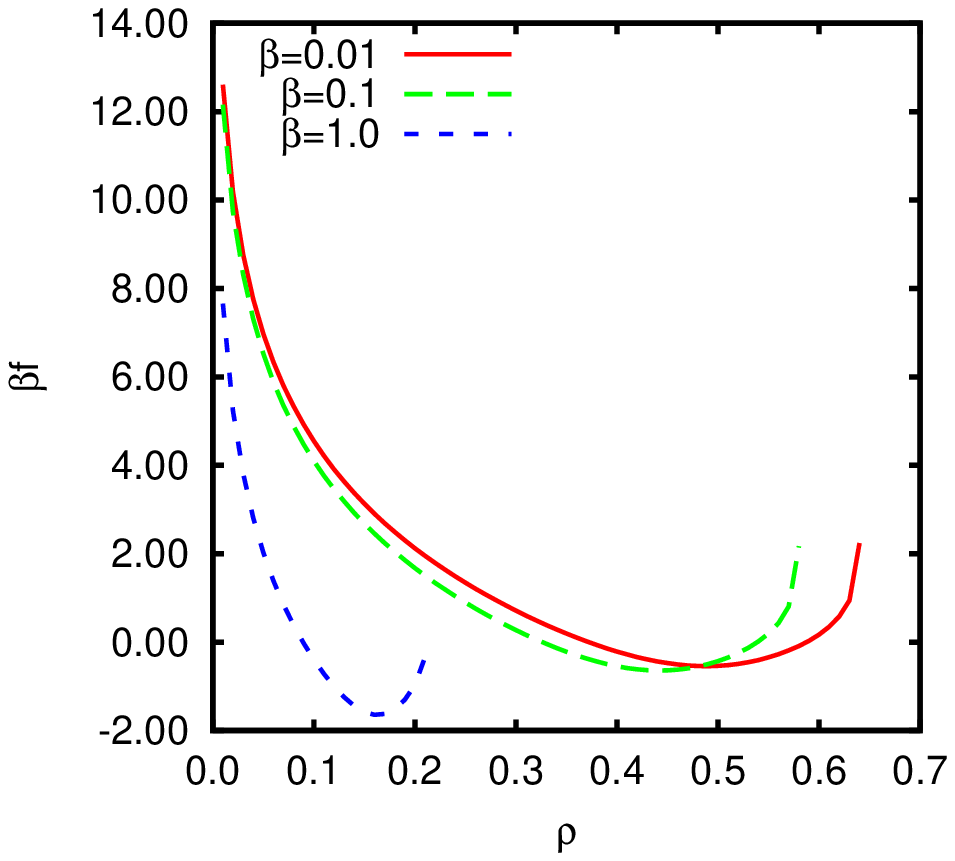}
\end{center}
\caption{(Color online) The upper panel shows the RPA validity region of the phase
diagram, in the Janus case, for
$z_1=1,z_{10}=e^{10\beta},z_{40}=e^{40\beta}$, in the 
absence of any excluded volume effect. At
$\beta=0$ the $C=0$ equation has solution $\rho\simeq
0.647933\ldots$. Note that the validity region is in the small density
region, where the contribution from the quadratic fluctuations of the
theory dominates, and the whole theory is expected to be less
significant. The lower panel shows the free energy density as a
function of the average density. The rapid increase at
high density is indicative of the limit of the RPA applicability.}    
\label{fig:valj}
\end{figure}

Choosing $z_1=1,z_{10}=e^{10\beta},$ and $z_{40}=e^{40\beta}$ we find
that at small densities (where the theory is expected to be not good)
the effective potential is repulsive (due to the quadratic
fluctuations in the theory) even without adding an excluded volume to
the polymer (see Section \ref{sec:effective}). The range of validity
of RPA in the phase diagram is shown in Fig. \ref{fig:valj}. 
In the same Figure we show the behavior of the free energy density which
clearly shows the signature for the breakdown of the RPA theory at
high density.

\begin{figure}[h!]
\begin{center}
\includegraphics[width=12cm]{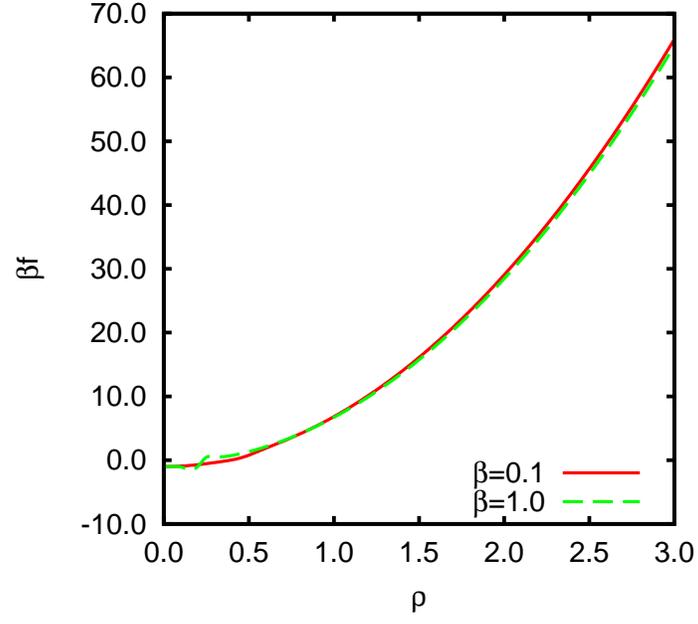}
\end{center}
\caption{(Color online) Shows the free energy density as a function of the average
density in the Janus case, for
$z_1=1,z_{10}=e^{10\beta},z_{40}=e^{40\beta}$, and $v=15$.} 
\label{fig:fvj}
\end{figure}

Notice that, also in this case, $\lim_{\bar{\rho}\to \infty} C=0$ so
that by adding a small excluded volume will allow to reach the high
densities domain with RPA. 
At a fixed temperature $C$, as a function of density, has a global
maximum, so that choosing the excluded volume $v$ bigger than this
value, the RPA can be made valid at any density (see
Fig. \ref{fig:fvj}). Moreover we expect the theory
to give consistent results in the high density regime. 

Our choice for the fugacities is justified a posteriori since for $\beta<1$
we are in the high temperature regime of the Janus vapor \cite{Sciortino2009}
where the internal energy of a cluster of $i$ Janus particles is with
a good approximation given by $-(i-1)\approx -i$ (corresponding to
a completely stretched cluster).

Since $Z_N$ is a grand-canonical in the clusters of Janus
particles, we can take derivatives with respect to the generalized
fugacity $z_i$ to determine the concentration $n_i$ of clusters 
of $i$ Janus particles as follows
\bq \label{ni}
n_i=\frac{i\langle N_i\rangle}{N}=
\frac{i}{N}\frac{\partial\ln Z_N}{\partial\ln z_i}~,
\eq 
where $\ln Z_N=V[A+\ln(-2C)/2]$.

\begin{figure}[h!]
\begin{center}
\includegraphics[width=12cm]{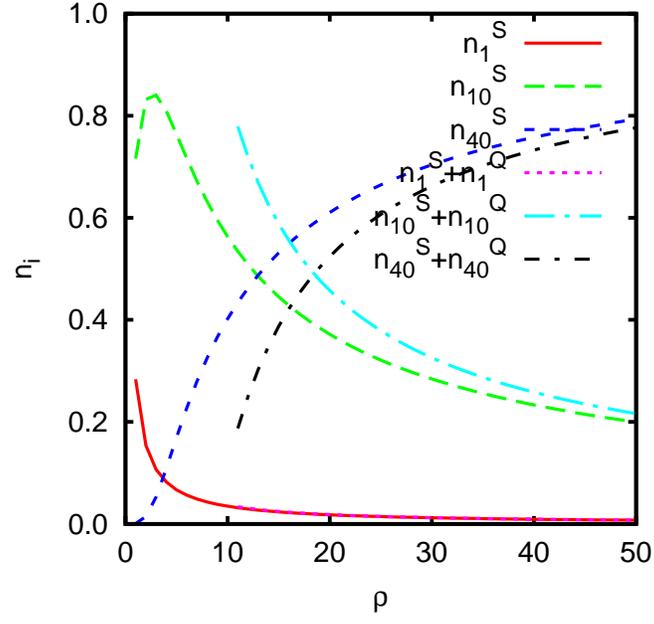}
\end{center}
\caption{(Color online) In the Janus case for
$z_1=1,z_{10}=e^{10\beta},z_{40}=e^{40\beta}$, $v=15$, and $\beta=1$ shows the
concentrations of clusters of 1, 10, and 40 particles as a function of
the density.}  
\label{fig:nj}
\end{figure}

\begin{figure}[h!]
\begin{center}
\includegraphics[width=12cm]{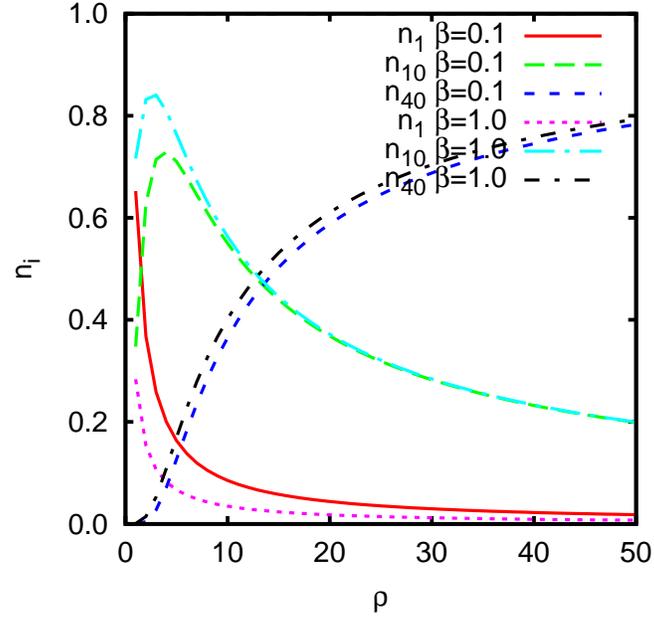}
\end{center}
\caption{(Color online) In the Janus case for
$z_1=1,z_{10}=e^{10\beta},z_{40}=e^{40\beta}$, and $v=15$, shows the
concentrations of clusters of 1 ($n_1^S$), 10 ($n_{10}^S$), and 40 ($n_{40}^S$)
particles as a function of the density when we do not use the
logarithmic correction in Eq. (\ref{M}).} 
\label{fig:njm}
\end{figure}

A graph of the concentrations as a function of the average polymer
density is shown in Fig. \ref{fig:nj}.
From Fig. \ref{fig:nj} one can see the difference between $n_x=n^S_x$ and
$n_x=n^S_x+n^Q_x$ for $x=1,10,$ and $40$ and $\beta=1$. Note that the
conservation of 
particles $n_1+n_{10}+n_{40}=1$ is exactly satisfied at all
densities and temperatures in both cases. In Fig. \ref{fig:njm} we
show the dependence of the concentrations from the temperature. 
We thus would say that at sufficiently high densities the
vesicles appear and as a consequence the micelles are reduced.

We conclude that this suggests strong dominance of non-clustering at
low densities.  As the density is increased smaller clusters and
eventually larger clusters dominate the linking behavior.   

\section{Conclusions}
\label{sec:conclusions}

In this work we have studied and developed a field-theoretical
formalism for a polymer immersed in an ideal mixture of clustering
centers.  These centers cause clustering of either $a$ particles or
$b$ particles, {\sl i.e.}  clusters of either species are
monodisperse.  The field-theory couples fields associated with
stickers to the polymer chain density and provides a formally exact
expression for the partition function (canonical in polymer and
grand-canonical for the clustering centers).  We showed that it is
possible to compute quantities using the nonlinear theory by means of
a saddle point approximation and we argue that the approximation
improves as the density of the polymer chain increases.  \bn{The
  current system and the choice of implementation of additional fields
  enabled us to derive saddle-point equations that are simpler than
  those that arise in some other formalisms by not requiring the solution
  of nonlinear integral equations.  The benefit of the local
  saddle-point equations is that they also enable a relatively simple
  analysis of the stability and applicability considerations of the
  theory.}

For a homogeneous, dense polymer system, we computed the effective
interaction potential (up to quadratic density fluctuations) and
computed properties of the structure factor within the random phase
approximation.  As expected, the addition of an excluded volume
interaction will compensate for the attraction due to aggregation
effects and extend the validity of the RPA.  The effective pairwise
potential obtained in this approximation has interesting, nontrivial
density dependence.  Another clear consequence of increasing chain
density is the growing dominance of the higher-functional clustering
centers.

The nature of the clustering process is definitely of importance in
aggregating polymer systems.  (Recently, a theory for cluster
formation in homopolymer melts was introduced by Semenov
\cite{Semenov2009}.)  One motivation for our study is the closed
multimerization scenario suggested by particles in a Janus fluid,
where micelles ($a=10$) and vesicles ($b=40$) are known to occur
\cite{Sciortino2009,Giacometti2009}. Future work will focus the
attention on the stability of such Janus-type multimers when connected
to a polymer with more detailed models of the cluster itself.

\appendix
\section{Coefficients $A,B,$ and $C$ for the $a=2,b=0$ case}
\label{app:AB1}
The $A,B,$ and $C$ coefficients are given by
\bq \nonumber
A_\pm&=&\frac{1}{8z_2}\left\{
-z_1^2\pm z_1\sqrt{z_1^2+8\bar{\rho}z_2}+8\bar{\rho}z_2
\ln\left(\frac{4\bar{\rho}z_2}{-z_1\pm\sqrt{z_1^2+8\bar{\rho}z_2}}\right)+
\right.\\
&&\left. 4z_2\left[\bar{\rho}+
\ln\left(1\pm\frac{z_1}{\sqrt{z_1^2+8\bar{\rho}z_2}}\right)-\ln 8
\right]\right\}~,\\
B_\pm&=&-\frac{2z_2}{z_1^2+8\bar{\rho}z_2}+\frac{1}{4\bar{\rho}}\left(
1\mp\frac{z_1}{\sqrt{z_1^2+8\bar{\rho}z_2}}\right)+
\ln\left(\frac{4\bar{\rho}z_2}{-z_1\pm\sqrt{z_1^2+8\bar{\rho}z_2}}\right)~,\\
C_\pm&=&\frac{8z_2^2}{(z_1^2+8\bar{\rho}z_2)^2}
+\frac{1}{8\bar{\rho}^2}\left[-1\pm\frac{z_1}{\sqrt{z_1^2+8\bar{\rho}z_2}}
+\bar{\rho}\left(2\mp
\frac{2z_1(z_1^2-2z_2+8\bar{\rho}z_2)}{(z_1^2+8\bar{\rho}z_2)^{3/2}}
\right)
\right]~.
\eq

\section{The Random Phase Approximation}
\label{app:RPA}
For the polymer chain, with a Kuhn length $\ell$, we can write
\bq \nonumber
&&\int\left\{\prod d\RR\right\}\left\{\prod G\right\}=\\ \nonumber
&&\int[d\rho]\int[d\zeta]\int[d\RR]\,e^{-\frac{3}{2\ell}\int_0^L ds\,\dot{\RR}^2(s)}
e^{i\int d\rr\,\zeta(\rr)[\rho(\rr)-\int_0^L
\frac{ds}{\ell}\,\delta(\rr-\RR(s))]}=\\ \nonumber
&&\int[d\rho]\int[d\zeta]\int[d\RR]\,e^{-\frac{3}{2\ell}\int_0^L ds\,\dot{\RR}^2(s)}
e^{i\int d\rr\,\zeta(\rr)\rho(\rr)}
\left[1-i\int d\rr\,\int_0^L
\frac{ds}{\ell}\,\zeta(\rr)\delta(\rr-\RR(s))-\right.\\
\label{RPA1}
&&\left.\frac{1}{2}\int d\rr\,\int_0^L 
\frac{ds}{\ell}\,\zeta(\rr)\delta(\rr-\RR(s))
\int d\rr^{'}\,\int_0^L \frac{ds^{'}}{\ell}\,\zeta(\rr^{'})\delta(\rr^{'}-\RR(s^{'}))+
\ldots\right]~,
\eq
where $L=N\ell$ is the total polymer length and the dot denotes a
derivative with respect to $s$.

Now the first term gives just a normalization constant ${\cal N}$. To
calculate the second term we introduce the polymer center of mass
$\RR_0$ so that $\RR(s)=\RR_0+\Delta \RR(s)$ and write
\bq \nonumber
&&\int[d\zeta]\int[d\RR]\,e^{-\frac{3}{2\ell}\int_0^L ds\,\dot{\RR}^2(s)}
e^{i\int d\rr\,\zeta(\rr)\rho(\rr)}
\int d\rr\int_0^L \frac{ds}{\ell}\,\zeta(\rr)\delta(\rr-\RR(s))=\\ \nonumber
&&\int[d\zeta]\int[d\RR]\,e^{-\frac{3}{2\ell}\int_0^L ds\,\dot{\RR}^2(s)}
e^{i\int d\rr\,\zeta(\rr)\rho(\rr)}
\int d\rr\int_0^L \frac{ds}{\ell}\,\zeta(\rr)\frac{1}{V}\sum_{\kk}
e^{i\kk\rr-i\kk\RR(s)}=\\ \nonumber
&&\int[d\zeta]\int[d\Delta\RR]\int d\RR_0\,
e^{-\frac{3}{2\ell}\int_0^L ds\,\Delta\dot{\RR}^2(s)}e^{i\int d\rr\,\zeta(\rr)\rho(\rr)}
\int d\rr\int_0^L \frac{ds}{\ell}\frac{1}{V}\sum_{\kk}\zeta(\rr)e^{i\kk\rr}
e^{-i\kk(\RR_0+\Delta\RR(s))}=\\ \nonumber
&&\int[d\zeta]e^{i\int d\rr\,\zeta(\rr)\rho(\rr)}
{\cal N}\frac{1}{V}\int d\rr\int_0^L \frac{ds}{\ell}\sum_{\kk}
\zeta(\rr)\delta_{\kk,0}=\\
&&\int[d\zeta]e^{i\int d\rr\,\zeta(\rr)\rho(\rr)}
{\cal N}\frac{N}{V}\int d\rr\,\zeta(\rr)~,
\eq
where $V$ is the volume of the box and $\delta_{\kk,0}$ is the
Kronecker delta. 

The third term gives
\bq \nonumber
&&\int[d\zeta]e^{i\int d\rr\,\zeta(\rr)\rho(\rr)}\int[d\Delta\RR]\int d\RR_0\,
e^{-\frac{3}{2\ell}\int_0^L ds\,\Delta\dot{\RR}^2(s)} \frac{1}{V^2}
\sum_{\kk}\sum_{\kk^\prime}\int d\rr\int_0^L\frac{ds}{\ell}\int
d\rr^{'}\int_0^L\frac{ds^{'}}{\ell}\times\\ \nonumber
&&\zeta(\rr)\zeta(\rr^{'})e^{i\kk\rr}e^{i\kk^{'}\rr^{'}}e^{i(\kk+\kk^{'})\RR_0}
e^{i\kk\Delta\RR(s)}e^{i\kk^{'}\Delta\RR(s^{'})}=\\
&&\int[d\zeta]e^{i\int d\rr\,\zeta(\rr)\rho(\rr)}{\cal N}\frac{1}{V^2}
\int_0^L\frac{ds}{\ell}\int_0^L\frac{ds^{'}}{\ell}\sum_{\kk}
\tilde{\zeta}(\kk)\tilde{\zeta}(-\kk)
\left\langle e^{i\kk(\RR(s)-\RR(s^{'}))}\right\rangle_0~,
\eq
where we denoted with the average 
\bq
\left\langle\ldots\right\rangle_0=\frac{\int[d\RR]
e^{-\frac{3}{2\ell}\int_0^L ds\,\dot{\RR}^2(s)}[\ldots]}
{\int[d\RR]e^{-\frac{3}{2\ell}\int_0^L ds\,\dot{\RR}^2(s)}}~,
\eq
and with the tilde the Fourier transform \cite{FT}.

Now the average $\left\langle
e^{i\kk(\RR(s)-\RR(s^{'}))}\right\rangle_0$ can be easily calculated
by discretizing the polymer and integrating over the bond vectors
$\bb_i=\RR_{i+1}-\RR_i$, as follows 
\bq
\left\langle e^{i\kk(\RR(s)-\RR(s^{'}))}\right\rangle_0=
\frac{\int\prod_i d\bb_ie^{-\frac{3}{2\ell^2}\sum_i\bb_i^2}e^{i\kk(\bb_1+\bb_2+\ldots+\bb_n)}}
{\int\prod_i d\bb_ie^{-\frac{3}{2\ell^2}\sum_i\bb_i^2}}
=\left(e^{-\frac{k^2\ell^2}{6}}\right)^n~,
\eq
where $n=|s-s^{'}|/\ell$. Since
\bq
\int_0^L ds\int_0^L ds^{'}\,
e^{-a|s-s^{'}|}=\frac{2(aL-1+e^{-aL})}{a^2}~,
\eq
we can introduce the function 
\bq \label{S(k)}
\tilde{S}_0(k)=\frac{72(k^2\ell L/6-1+e^{-\frac{k^2\ell L}{6}})}{V^2k^4\ell^4}~,
\eq
with $\tilde{S}_0(0)=(N/V)^2=\bar{\rho}^2$.

Then the expression we started with in Eq. (\ref{RPA1}) can be
rewritten, omitting the functional integral over the density
collective variable, as
\bq
{\cal N}\int[d\tilde{\zeta}]e^{\frac{i}{V}\sum_{\kk}\tilde{\zeta}(\kk)\tilde{\rho}(-\kk)}
\left[1-i\frac{N}{V}\tilde{\zeta}(0)-\frac{1}{2V}
\sum_{\kk}\tilde{\zeta}(\kk)V\tilde{S}_0(k)
\tilde{\zeta}(-\kk)+
\ldots\right]~.
\eq 
We can now reconstruct the exponential to obtain,
\bq \nonumber
\int[d\tilde{\zeta}]e^{\frac{i}{V}\sum_{\kk}\tilde{\zeta}(\kk)\tilde{\rho}(-\kk)}
e^{-\frac{1}{2V}\sum_{\kk}\tilde{\zeta}(\kk)V\tilde{S}_0(k)\tilde{\zeta}(-\kk)
-i\frac{N}{V}\tilde{\zeta}(0)}=\\
{\cal N}^\prime
e^{-\frac{1}{2V}\sum_{\kk}\ln(\tilde{S}_0(k)V)}
e^{-\frac{1}{2V}\sum_{\kk}\Delta\tilde{\rho}(\kk)
\frac{\tilde{S}_0^{-1}(k)}{V}\Delta\tilde{\rho}(-\kk)}~.
\eq
Here
$\Delta\tilde{\rho}(\kk)=\tilde{\rho}(\kk)-\bar{\rho}V\delta_{\kk,0}$.

\an{
\section{The Gaussian distribution}
\label{app:GD}
The Gaussian distribution function for a set of real variables
$x_1,x_2,\ldots,x_N$ is defined as 
\bq \label{df}
\Psi(x_1,x_2,\ldots,x_N)=C\exp\left[-\frac{1}{2}\sum_{n,m}A_{nm}x_nx_m\right]~,
\eq
where $A_{nm}$ is a symmetric positive definite matrix and $C$ is a
normalization constant given by the requirement
$\int_{-\infty}^{\infty}\cdots\int_{-\infty}^{\infty}\prod_ndx_n\Psi=1$. 

Let $\langle\ldots\rangle$ be the average of the distribution function
of Eq. (\ref{df}),
\bq
\langle\ldots\rangle=\int_{-\infty}^{\infty}\cdots\int_{-\infty}^{\infty}
\prod_ndx_n\ldots\Psi(x_1,x_2,\ldots,x_N)~,
\eq
then, it can be proved \cite{DoiEdwards}, that
\bq \label{xnxm}
\langle x_nx_m\rangle=[A^{-1}]_{nm}~.
\eq 

In general we have the following formula (Wick's theorem)
\bq
\langle x_{n_1}x_{n_2}\ldots x_{n_{2p}}\rangle=
\sum_{\text{all pairing}}\langle x_{m_1}x_{m_2}\rangle\langle x_{m_3}x_{m_4}\rangle
\ldots\langle x_{m_{2p-1}}x_{m_{2p}}\rangle~.
\eq

If the subscript $n$ of $x_n$ is regarded as a continuous variable,
the set of points $(x_1,x_2,\ldots,x_N)$ represents a continuous
function, and the integral over the set $(x_1,x_2,\ldots,x_N)$ reduces
to the integration over all the function, and it is called the {\sl
  functional integral}. It is denoted by the symbol $[dx]$, {\sl i.e.}
$\int\prod_ndx_n\ldots\rightarrow\int[dx]\ldots$. 

Consider now the following Gaussian distribution functional
\bq
\Psi[\phi]=C\exp\left[-\frac{1}{2}\int_{-\infty}^{\infty}dx
\phi^2(x)\right]~,
\eq
where $\phi$ is a real function, then using the continuous limit of
Eq. (\ref{xnxm}) we find 
\bq \label{phiphi}
\langle\phi(x)\phi(x^\prime)\rangle=\delta(x-x^\prime)~.
\eq 
where $\delta$ is the Dirac delta function.

If now $\phi=\phi_1+i\phi_2$ is a complex function we consider the
Gaussian distribution functional
\bq
\Psi[\phi,\phi^\star]=C\exp\left[-\int_{-\infty}^{\infty}dx
\phi(x)\phi^\star(x)\right]~.
\eq
Now we find from Eq. (\ref{phiphi})
\bq \nonumber
\langle\phi(x)\phi^\star(x^\prime)\rangle&=&
\langle\phi_1(x)\phi_1(x^\prime)+\phi_2(x)\phi_2(x^\prime)\rangle=\\
&&\frac{1}{2}\delta(x-x^\prime)+\frac{1}{2}\delta(x-x^\prime)=
\delta(x-x^\prime)~,
\eq 
and
\bq \nonumber
\langle\phi(x)\phi(x^\prime)\rangle&=&
\langle\phi_1(x)\phi_1(x^\prime)-\phi_2(x)\phi_2(x^\prime)
+i\phi_1(x)\phi_2(x^\prime)+i\phi_2(x)\phi_1(x^\prime)\rangle=\\
&&\frac{1}{2}\delta(x-x^\prime)-\frac{1}{2}\delta(x-x^\prime)=
0~.
\eq 
}

{
\section{Field theory without 1-clusters}
\label{app:alternativefieldtheory}

An alternative way to formulate the clustering, without the 
use of clusters of size 1 is presented below.  We shall show that
a simple mapping reduces again to a special case of 
Eq.~(\ref{eq:fieldtheory}).

Consider a system in which we have only clusters of sizes $a$ and $b$
but no ``inert'' clusters of size 1.  As explained in Section~\ref{sec:model}
the functional integration over the fields $\varphi$ and $\varphi^\star$ requires
matching each $\varphi$ of a clustering center with a $\varphi^\star$ on the 
polymer, permitting no unmatched $\varphi$ and $\varphi^\star$ pairs.  Since size
$a$ and $b$ clusters do not necessarily attach to each potential
site on the polymer, all possible attachment sites have to be generated.
The product
\begin{displaymath}
\prod_{i=1}^N \left( 1 + \varphi^\star (\bm{R}_i) \right)
\end{displaymath}
produces all equally weighted possibilities of the attching to the
the sites $\left\{ \bm{R}_i \right\}, \forall i \in \{1, \ldots, N\}$
of a given polymer configuration.

The analog to Eq.~(\ref{fieldtheory0}) then becomes
\begin{eqnarray}
{\cal Z}'_{N_a,N_b}&=&{\cal N}\int d\RR_1\ldots d\RR_N
\an{e^{-v\sum_{n,m=1}^N\delta(\RR_n-\RR_m)}}
\int[d\varphi][d\varphi^\star]\,
e^{-\int d\rr\,\varphi(\rr)\varphi^\star(\rr)}\times\\ \nonumber
&&\left( 1+ \varphi^\star(\RR_1)\right) 
G(\RR_1,\RR_2)\left( 1+ \varphi^\star(\RR_2) \right) G(\RR_2,\RR_3)\ldots
G(\RR_{N-1},\RR_N)\left( 1+\varphi^\star(\RR_N)\right)\times\\
&& \frac{1}{N_a!}\left(\int d\rr\,z_a\varphi^a(\rr)\right)^{N_a}
\frac{1}{N_b!}\left(\int d\rr\,z_b\varphi^b(\rr)\right)^{N_b},
\end{eqnarray}
leading by the same procedure as described in Section~\ref{sec:model}
to analog of Eq.~(\ref{eq:fieldtheory})
\bq \nonumber
Z'_N&=&{\cal N}\int[d\varphi][d\varphi^\star]\left\{\prod d\RR\right\}
\left\{\prod G\right\}d^N 
\exp\left(-\int d\rr\, \varphi(\rr)\varphi^\star(\rr)+\right.\\
&&\left.\int d\rr\,\rho(\rr)\ln(1+\varphi^\star(\rr)/d)+
z_a\int d\rr\,\varphi^a(\rr)+
z_b\int d\rr\,\varphi^b(\rr)\right).
\label{eq:fieldtheory2}
\eq
We see that the trivial transformation $\varphi^\star \rightarrow
\varphi^\star -1$  
in Eq.~(\ref{eq:fieldtheory2}) above leads to the orgininal field-theoretic
equation in the main text, Eq.~(\ref{eq:fieldtheory}), with $z_1 \rightarrow 1$.
For this reason we treat the marginally more general case in this paper.
}

\begin{acknowledgments}
R.F. gratefully acknowledges support from the NITheP of South Africa.
K.K.M.-N. gratefully acknowledges the support of the National Research
Foundation. 
\end{acknowledgments}

\begin{thebibliography}{53}
\expandafter\ifx\csname natexlab\endcsname\relax\def\natexlab#1{#1}\fi
\expandafter\ifx\csname bibnamefont\endcsname\relax
  \def\bibnamefont#1{#1}\fi
\expandafter\ifx\csname bibfnamefont\endcsname\relax
  \def\bibfnamefont#1{#1}\fi
\expandafter\ifx\csname citenamefont\endcsname\relax
  \def\citenamefont#1{#1}\fi
\expandafter\ifx\csname url\endcsname\relax
  \def\url#1{\texttt{#1}}\fi
\expandafter\ifx\csname urlprefix\endcsname\relax\def\urlprefix{URL }\fi
\providecommand{\bibinfo}[2]{#2}
\providecommand{\eprint}[2][]{\url{#2}}

\bibitem[{\citenamefont{Hoy and Fredrickson}(2009)}]{HoyFredrickson2009}
\bibinfo{author}{\bibfnamefont{R.~S.} \bibnamefont{Hoy}} \bibnamefont{and}
  \bibinfo{author}{\bibfnamefont{G.~H.} \bibnamefont{Fredrickson}},
  \bibinfo{journal}{J. Chem. Phys.} \textbf{\bibinfo{volume}{131}},
  \bibinfo{pages}{224902} (\bibinfo{year}{2009}).

\bibitem[{\citenamefont{Rubinstein and
  Dobrynin}(1999)}]{RubinsteinDobrynin1999}
\bibinfo{author}{\bibfnamefont{M.}~\bibnamefont{Rubinstein}} \bibnamefont{and}
  \bibinfo{author}{\bibfnamefont{A.~V.} \bibnamefont{Dobrynin}},
  \bibinfo{journal}{Current Opinion in Colloid {\&} Interface Science}
  \textbf{\bibinfo{volume}{4}}, \bibinfo{pages}{83} (\bibinfo{year}{1999}).

\bibitem[{\citenamefont{Semenov and Rubinstein}(1998)}]{SemenovRubinstein1998}
\bibinfo{author}{\bibfnamefont{A.~N.} \bibnamefont{Semenov}} \bibnamefont{and}
  \bibinfo{author}{\bibfnamefont{M.}~\bibnamefont{Rubinstein}},
  \bibinfo{journal}{Macromol.} \textbf{\bibinfo{volume}{31}},
  \bibinfo{pages}{1373} (\bibinfo{year}{1998}).

\bibitem[{\citenamefont{Muthukumar}(1996)}]{Muthukumar1996}
\bibinfo{author}{\bibfnamefont{M.}~\bibnamefont{Muthukumar}},
  \bibinfo{journal}{J. Chem. Phys.} \textbf{\bibinfo{volume}{104}},
  \bibinfo{pages}{691} (\bibinfo{year}{1996}).

\bibitem[{\citenamefont{Nyrkova and Semenov}(2005)}]{NyrkovaSemenov2005}
\bibinfo{author}{\bibfnamefont{I.~A.} \bibnamefont{Nyrkova}} \bibnamefont{and}
  \bibinfo{author}{\bibfnamefont{A.~N.} \bibnamefont{Semenov}},
  \bibinfo{journal}{Eur. Phys. J. E} \textbf{\bibinfo{volume}{17}},
  \bibinfo{pages}{327} (\bibinfo{year}{2005}).

\bibitem[{\citenamefont{{R. T. Deam and S. F. Edwards}}(1976)}]{Deam1976}
\bibinfo{author}{\bibnamefont{{R. T. Deam and S. F. Edwards}}},
  \bibinfo{journal}{Phil. Trans. R. Soc. London A. Math. Phys. Sciences}
  \textbf{\bibinfo{volume}{280}}, \bibinfo{pages}{317} (\bibinfo{year}{1976}).

\bibitem[{\citenamefont{{S. I. Kuchanov, S. V. Korolev, and S. V.
  Panyukov}}(1988)}]{Kuchanov1988}
\bibinfo{author}{\bibnamefont{{S. I. Kuchanov, S. V. Korolev, and S. V.
  Panyukov}}}, \bibinfo{journal}{Adv. Chem. Phys.}
  \textbf{\bibinfo{volume}{72}}, \bibinfo{pages}{115} (\bibinfo{year}{1988}).

\bibitem[{\citenamefont{{A. V. Ermoshkin and I. Y.
  Erukhimovich}}(1999)}]{Ermoshkin1999b}
\bibinfo{author}{\bibnamefont{{A. V. Ermoshkin and I. Y. Erukhimovich}}},
  \bibinfo{journal}{J. Chem. Phys.} \textbf{\bibinfo{volume}{110}},
  \bibinfo{pages}{1781} (\bibinfo{year}{1999}).

\bibitem[{\citenamefont{{S. Kuchanov, H. Slot, and A.
  Stroeks}}(2004)}]{Kuchanov2004}
\bibinfo{author}{\bibnamefont{{S. Kuchanov, H. Slot, and A. Stroeks}}},
  \bibinfo{journal}{Prog. Polym. Sci.} \textbf{\bibinfo{volume}{29}},
  \bibinfo{pages}{563} (\bibinfo{year}{2004}).

\bibitem[{\citenamefont{{A. Kudlay and I. Y. Erukhimovich}}(2001)}]{Kudlay2001}
\bibinfo{author}{\bibnamefont{{A. Kudlay and I. Y. Erukhimovich}}},
  \bibinfo{journal}{Macromol. Theory Simul.} \textbf{\bibinfo{volume}{10}},
  \bibinfo{pages}{542} (\bibinfo{year}{2001}).

\bibitem[{\citenamefont{{F. Sciortino, E. Bianchi, J. F. Douglas, and P.
  Tartaglia}}(2007)}]{Sciortino2007}
\bibinfo{author}{\bibnamefont{{F. Sciortino, E. Bianchi, J. F. Douglas, and P.
  Tartaglia}}}, \bibinfo{journal}{J. Chem. Phys.}
  \textbf{\bibinfo{volume}{126}}, \bibinfo{pages}{194903}
  (\bibinfo{year}{2007}).

\bibitem[{\citenamefont{{Yu. V. Kalyuzhnyi, C. -T. Lin, and G.
  Stell}}(1998)}]{Kalyuzhnyi1998}
\bibinfo{author}{\bibnamefont{{Yu. V. Kalyuzhnyi, C. -T. Lin, and G. Stell}}},
  \bibinfo{journal}{J. Chem. Phys.} \textbf{\bibinfo{volume}{108}},
  \bibinfo{pages}{6525} (\bibinfo{year}{1998}).

\bibitem[{\citenamefont{{R. Nagarajan}}(1989)}]{Nagarajan1989}
\bibinfo{author}{\bibnamefont{{R. Nagarajan}}}, \bibinfo{journal}{J. Chem.
  Phys.} \textbf{\bibinfo{volume}{90}}, \bibinfo{pages}{1980}
  (\bibinfo{year}{1989}).

\bibitem[{\citenamefont{{F. Ganazzoli, G. Raos, and
  G.Allegra}}(1999)}]{Ganazzoli1999}
\bibinfo{author}{\bibnamefont{{F. Ganazzoli, G. Raos, and G.Allegra}}},
  \bibinfo{journal}{Macromol. Theory Simul.} \textbf{\bibinfo{volume}{8}},
  \bibinfo{pages}{65} (\bibinfo{year}{1999}).

\bibitem[{\citenamefont{{Tanaka}}(1990)}]{Tanaka1990}
\bibinfo{author}{\bibnamefont{{Tanaka}}}, \bibinfo{journal}{Macromolecules}
  \textbf{\bibinfo{volume}{23}}, \bibinfo{pages}{3784} (\bibinfo{year}{1990}).

\bibitem[{\citenamefont{{B. Xu, A. Yekta, L. Li, Z. Masoumi, and M. A.
  Winnik}}(1996)}]{Xu1996}
\bibinfo{author}{\bibnamefont{{B. Xu, A. Yekta, L. Li, Z. Masoumi, and M. A.
  Winnik}}}, \bibinfo{journal}{Colloid and Surfaces A}
  \textbf{\bibinfo{volume}{112}}, \bibinfo{pages}{239} (\bibinfo{year}{1996}).

\bibitem[{\citenamefont{Semenov}(2009)}]{Semenov2009}
\bibinfo{author}{\bibfnamefont{A.~N.} \bibnamefont{Semenov}},
  \bibinfo{journal}{Macromol.} \textbf{\bibinfo{volume}{42}},
  \bibinfo{pages}{6761} (\bibinfo{year}{2009}).

\bibitem[{\citenamefont{{S. M. Loverde, A. V. Ermoshkin, and M. Olvera de la
  Cruz}}(2005)}]{Loverde2005}
\bibinfo{author}{\bibnamefont{{S. M. Loverde, A. V. Ermoshkin, and M. Olvera de
  la Cruz}}}, \bibinfo{journal}{J. Polym. Sci. Part B: Polym. Physics}
  \textbf{\bibinfo{volume}{43}}, \bibinfo{pages}{796} (\bibinfo{year}{2005}).

\bibitem[{\citenamefont{{I. Y. Erukhimovich and A. V.
  Ermoshkin}}(1999)}]{Erukhimovich1999}
\bibinfo{author}{\bibnamefont{{I. Y. Erukhimovich and A. V. Ermoshkin}}},
  \bibinfo{journal}{JETP} \textbf{\bibinfo{volume}{88}}, \bibinfo{pages}{538}
  (\bibinfo{year}{1999}).

\bibitem[{\citenamefont{{F. Sciortino, A. Giacometti and G.
  Pastore}}(2009)}]{Sciortino2009}
\bibinfo{author}{\bibnamefont{{F. Sciortino, A. Giacometti and G. Pastore}}},
  \bibinfo{journal}{Phys. Rev. Lett.} \textbf{\bibinfo{volume}{103}},
  \bibinfo{pages}{237801} (\bibinfo{year}{2009}).

\bibitem[{\citenamefont{{Achille Giacometti, Fred Lado, Julio Largo, Giorgio
  Pastore, and Francesco Sciortino}}(2009)}]{Giacometti2009}
\bibinfo{author}{\bibnamefont{{Achille Giacometti, Fred Lado, Julio Largo,
  Giorgio Pastore, and Francesco Sciortino}}}, \bibinfo{journal}{J. Chem.
  Phys.} \textbf{\bibinfo{volume}{131}}, \bibinfo{pages}{174114}
  (\bibinfo{year}{2009}).

\bibitem[{\citenamefont{{R. Fantoni , A. Giacometti, F. Sciortino, and G.
  Pastore}}(2011)}]{Fantoni2011}
\bibinfo{author}{\bibnamefont{{R. Fantoni , A. Giacometti, F. Sciortino, and G.
  Pastore}}}, \bibinfo{journal}{Soft Matter} \textbf{\bibinfo{volume}{7}},
  \bibinfo{pages}{2419} (\bibinfo{year}{2011}).

\bibitem[{\citenamefont{Edwards}(1988)}]{Edwards1988}
\bibinfo{author}{\bibfnamefont{S.~F.} \bibnamefont{Edwards}},
  \bibinfo{journal}{J. Phys. France} \textbf{\bibinfo{volume}{49}},
  \bibinfo{pages}{1673} (\bibinfo{year}{1988}).

\bibitem[{\citenamefont{{S. F. Edwards and K. F.
  Freed}}(1970{\natexlab{a}})}]{Edwards1970a}
\bibinfo{author}{\bibnamefont{{S. F. Edwards and K. F. Freed}}},
  \bibinfo{journal}{J. Phys. C: Solid State Phys.}
  \textbf{\bibinfo{volume}{3}}, \bibinfo{pages}{739}
  (\bibinfo{year}{1970}{\natexlab{a}}).

\bibitem[{\citenamefont{{S. F. Edwards and K. F.
  Freed}}(1970{\natexlab{b}})}]{Edwards1970b}
\bibinfo{author}{\bibnamefont{{S. F. Edwards and K. F. Freed}}},
  \bibinfo{journal}{J. Phys. C: Solid State Phys.}
  \textbf{\bibinfo{volume}{3}}, \bibinfo{pages}{750}
  (\bibinfo{year}{1970}{\natexlab{b}}).

\bibitem[{\citenamefont{Gordon}(1962)}]{Gordon1962}
\bibinfo{author}{\bibfnamefont{M.}~\bibnamefont{Gordon}},
  \bibinfo{journal}{Proc. R. Soc. Lond. A} \textbf{\bibinfo{volume}{268}},
  \bibinfo{pages}{240} (\bibinfo{year}{1962}).

\bibitem[{\citenamefont{Gordon and Scantlebury}(1964)}]{Gordon1964}
\bibinfo{author}{\bibfnamefont{M.}~\bibnamefont{Gordon}} \bibnamefont{and}
  \bibinfo{author}{\bibfnamefont{G.~R.} \bibnamefont{Scantlebury}},
  \bibinfo{journal}{Trans. Faraday Soc.} \textbf{\bibinfo{volume}{60}},
  \bibinfo{pages}{604} (\bibinfo{year}{1964}).

\bibitem[{\citenamefont{Mohan et~al.}(2010)\citenamefont{Mohan, Elliot, and
  Fredrickson}}]{Mohan2010}
\bibinfo{author}{\bibfnamefont{A.}~\bibnamefont{Mohan}},
  \bibinfo{author}{\bibfnamefont{R.}~\bibnamefont{Elliot}}, \bibnamefont{and}
  \bibinfo{author}{\bibfnamefont{G.~H.} \bibnamefont{Fredrickson}},
  \bibinfo{journal}{J. Chem. Phys.} \textbf{\bibinfo{volume}{133}},
  \bibinfo{pages}{174903} (\bibinfo{year}{2010}).

\bibitem[{\citenamefont{Bohbot-Raviv et~al.}(2004)\citenamefont{Bohbot-Raviv,
  Snyder, and Wang}}]{BohbotRaviv2004}
\bibinfo{author}{\bibfnamefont{Y.}~\bibnamefont{Bohbot-Raviv}},
  \bibinfo{author}{\bibfnamefont{T.~M.} \bibnamefont{Snyder}},
  \bibnamefont{and} \bibinfo{author}{\bibfnamefont{Z.-G.} \bibnamefont{Wang}},
  \bibinfo{journal}{Langmuir} \textbf{\bibinfo{volume}{20}},
  \bibinfo{pages}{7860} (\bibinfo{year}{2004}).

\bibitem[{\citenamefont{Cates and Witten}(1986)}]{Cates1986}
\bibinfo{author}{\bibfnamefont{M.~E.} \bibnamefont{Cates}} \bibnamefont{and}
  \bibinfo{author}{\bibfnamefont{T.~A.} \bibnamefont{Witten}},
  \bibinfo{journal}{Macromol.} \textbf{\bibinfo{volume}{19}},
  \bibinfo{pages}{732} (\bibinfo{year}{1986}).

\bibitem[{\citenamefont{Nakamura and Shi}(2010)}]{Nakamura2010}
\bibinfo{author}{\bibfnamefont{I.}~\bibnamefont{Nakamura}} \bibnamefont{and}
  \bibinfo{author}{\bibfnamefont{A.-C.} \bibnamefont{Shi}},
  \bibinfo{journal}{J. Chem. Phys.} \textbf{\bibinfo{volume}{132}},
  \bibinfo{pages}{194103} (\bibinfo{year}{2010}).

\bibitem[{\citenamefont{Ermoshkin and {Olvera de la
  Cruz}}(2004)}]{Ermoshkin2004}
\bibinfo{author}{\bibfnamefont{A.~V.} \bibnamefont{Ermoshkin}}
  \bibnamefont{and} \bibinfo{author}{\bibfnamefont{M.}~\bibnamefont{{Olvera de
  la Cruz}}}, \bibinfo{journal}{J. Polym. Sc.: Part B: Polym. Phys.}
  \textbf{\bibinfo{volume}{42}}, \bibinfo{pages}{766} (\bibinfo{year}{2004}).

\bibitem[{\citenamefont{{K. M. Hong and J. Noolandi}}(1981)}]{Hong1981}
\bibinfo{author}{\bibnamefont{{K. M. Hong and J. Noolandi}}},
  \bibinfo{journal}{Macromolecules} \textbf{\bibinfo{volume}{14}},
  \bibinfo{pages}{727} (\bibinfo{year}{1981}).

\bibitem[{\citenamefont{{A. C. Shi, J. Noolandi, and R. C.
  Desai}}(1996)}]{Shi1996}
\bibinfo{author}{\bibnamefont{{A. C. Shi, J. Noolandi, and R. C. Desai}}},
  \bibinfo{journal}{Macromolecules} \textbf{\bibinfo{volume}{29}},
  \bibinfo{pages}{6487} (\bibinfo{year}{1996}).

\bibitem[{\citenamefont{Leibler}(1980)}]{Leibler1980}
\bibinfo{author}{\bibfnamefont{L.}~\bibnamefont{Leibler}},
  \bibinfo{journal}{Macromol.} \textbf{\bibinfo{volume}{13}},
  \bibinfo{pages}{1602} (\bibinfo{year}{1980}).

\bibitem[{\citenamefont{{V. N. Manoharan, M. T. Elsesser, and D. J.
  Pine}}(2003)}]{Manoharan2003}
\bibinfo{author}{\bibnamefont{{V. N. Manoharan, M. T. Elsesser, and D. J.
  Pine}}}, \bibinfo{journal}{Science} \textbf{\bibinfo{volume}{301}},
  \bibinfo{pages}{483} (\bibinfo{year}{2003}).

\bibitem[{\citenamefont{{A. B. Pawar and I. Kretzschmar}}(2010)}]{Pawar2010}
\bibinfo{author}{\bibnamefont{{A. B. Pawar and I. Kretzschmar}}},
  \bibinfo{journal}{Macromol. Rapid Commun.} \textbf{\bibinfo{volume}{31}},
  \bibinfo{pages}{150} (\bibinfo{year}{2010}).

\bibitem[{\citenamefont{{S. C. Glotzer and M. J. Solomon}}(2007)}]{Glotzer2007}
\bibinfo{author}{\bibnamefont{{S. C. Glotzer and M. J. Solomon}}},
  \bibinfo{journal}{Nature Materials} \textbf{\bibinfo{volume}{6}},
  \bibinfo{pages}{557} (\bibinfo{year}{2007}).

\bibitem[{\citenamefont{{Z. Zhang and S. C. Glotzer}}(2004)}]{Zhang2004}
\bibinfo{author}{\bibnamefont{{Z. Zhang and S. C. Glotzer}}},
  \bibinfo{journal}{Nano Letters} \textbf{\bibinfo{volume}{4}},
  \bibinfo{pages}{1407} (\bibinfo{year}{2004}).

\bibitem[{\citenamefont{{P. G. de Gennes}}(1992)}]{deGennes1992}
\bibinfo{author}{\bibnamefont{{P. G. de Gennes}}}, \bibinfo{journal}{Rev. Mod.
  Phys.} \textbf{\bibinfo{volume}{64}}, \bibinfo{pages}{645}
  (\bibinfo{year}{1992}).

\bibitem[{\citenamefont{{C. Casagrande, P. Fabre, M. Veyssi\'e, and E.
  Rapha\"el}}(1989)}]{Casagrande1989}
\bibinfo{author}{\bibnamefont{{C. Casagrande, P. Fabre, M. Veyssi\'e, and E.
  Rapha\"el}}}, \bibinfo{journal}{Europhys. Lett.}
  \textbf{\bibinfo{volume}{9}}, \bibinfo{pages}{251} (\bibinfo{year}{1989}).

\bibitem[{\citenamefont{{L. Hong, A. Cacciuto, E. Luijten, and S.
  Granick}}(2006)}]{Hong2006}
\bibinfo{author}{\bibnamefont{{L. Hong, A. Cacciuto, E. Luijten, and S.
  Granick}}}, \bibinfo{journal}{Nano Letters} \textbf{\bibinfo{volume}{6}},
  \bibinfo{pages}{2510} (\bibinfo{year}{2006}).

\bibitem[{\citenamefont{{A. Walther and A. H. M\"uller}}(2008)}]{Walther2008}
\bibinfo{author}{\bibnamefont{{A. Walther and A. H. M\"uller}}},
  \bibinfo{journal}{Soft Matter} \textbf{\bibinfo{volume}{4}},
  \bibinfo{pages}{663} (\bibinfo{year}{2008}).

\bibitem[{\citenamefont{{Y. Ding, H. C. \"Ottinger, A. D. Schl\"uter, and M.
  Kr\"oger}}(2007)}]{Ding2007}
\bibinfo{author}{\bibnamefont{{Y. Ding, H. C. \"Ottinger, A. D. Schl\"uter, and
  M. Kr\"oger}}}, \bibinfo{journal}{J. Chem. Phys.}
  \textbf{\bibinfo{volume}{127}}, \bibinfo{pages}{094904}
  (\bibinfo{year}{2007}).

\bibitem[{\citenamefont{{J. U. Kim and M. W. Matsen}}(2009)}]{Kim2009}
\bibinfo{author}{\bibnamefont{{J. U. Kim and M. W. Matsen}}},
  \bibinfo{journal}{Phys. Rev. Lett.} \textbf{\bibinfo{volume}{102}},
  \bibinfo{pages}{078303} (\bibinfo{year}{2009}).

\bibitem[{\citenamefont{{P. Haronska and T. A. Vilgis}}(1994)}]{Haronska1994a}
\bibinfo{author}{\bibnamefont{{P. Haronska and T. A. Vilgis}}},
  \bibinfo{journal}{Phys. Rev. E} \textbf{\bibinfo{volume}{50}},
  \bibinfo{pages}{325} (\bibinfo{year}{1994}).

\bibitem[{\citenamefont{{T. A. Vilgis and P. Haronska}}(1994)}]{Vilgis1994a}
\bibinfo{author}{\bibnamefont{{T. A. Vilgis and P. Haronska}}},
  \bibinfo{journal}{Macromol.} \textbf{\bibinfo{volume}{27}},
  \bibinfo{pages}{6465} (\bibinfo{year}{1994}).

\bibitem[{FT()}]{FT}
\bibinfo{note}{Notice that since the system has a finite volume $V$ the Fourier
  transform are of the discrete type. In the thermodynamic limit $V\to\infty$
  and $V^{-1}\sum_{\kk}\ldots\to\int d\kk/(2\pi)^3\ldots$.}

\bibitem[{\citenamefont{{K. K. M\"uller-Nedebock, S. F. Edwards, and T. C. B.
  McLeish}}(1999)}]{Nedebock1999}
\bibinfo{author}{\bibnamefont{{K. K. M\"uller-Nedebock, S. F. Edwards, and T.
  C. B. McLeish}}}, \bibinfo{journal}{J. Chem. Phys.}
  \textbf{\bibinfo{volume}{111}}, \bibinfo{pages}{8196} (\bibinfo{year}{1999}).

\bibitem[{\citenamefont{{M. Doi and S. F. Edwards}}(1986)}]{DoiEdwards}
\bibinfo{author}{\bibnamefont{{M. Doi and S. F. Edwards}}},
  \emph{\bibinfo{title}{The theory of polymer dynamics}}
  (\bibinfo{publisher}{Clarendon Press}, \bibinfo{address}{Oxford},
  \bibinfo{year}{1986}).

\bibitem[{\citenamefont{Doi}(1992)}]{Doi}
\bibinfo{author}{\bibfnamefont{M.}~\bibnamefont{Doi}},
  \emph{\bibinfo{title}{Introduction to Polymer Physics}}
  (\bibinfo{publisher}{Clarendon Press}, \bibinfo{address}{Oxford},
  \bibinfo{year}{1992}).

\bibitem[{\citenamefont{{T. Shimada, M. Doi, and K.
  Okano}}(1988)}]{Shimada1988}
\bibinfo{author}{\bibnamefont{{T. Shimada, M. Doi, and K. Okano}}},
  \bibinfo{journal}{J. Chem. Phys.} \textbf{\bibinfo{volume}{88}},
  \bibinfo{pages}{2815} (\bibinfo{year}{1988}).

\bibitem[{pcs()}]{pcsf}
\bibinfo{note}{Note that for $N_p$ polymer chains this expression should be
  modified as $\tilde{g}(k)=N_p/[\tilde{S}_0^{-1}(k)/V-2CN_p]/\bar{\rho}$}.

\end{thebibliography}

\clearpage
\clearpage
\clearpage
\clearpage
\clearpage
\clearpage
\clearpage
\clearpage
\clearpage
\clearpage
{\Large \bf Figures captions}\\\\\\
{\bf Figure 1:} A schematic representation of the role of the
field-theory. The fields $\varphi$ and $\varphi^\star$ are depicted as 
functions of spatial variables. Multiplication by
$\exp(-\int\varphi\varphi^\star)$ and subsequent functional
integration enforces the linking of the spatial coordinates 
between pairs of $\varphi$ and $\varphi^\star$
(in all possible ways).\\
\\
{\bf Figure 2:} Shows the polymer made up of $N$ equispaced links.\\
\\
{\bf Figure 3:} Shows the clusters of Janus particles: the micelles
are made of $a=10$ links whereas the vesicles of $b=40$ links.\\
\\
{\bf Figure 4:} (Color online) Shows the RPA validity region of the phase diagram, in the
$a=2,b=0$ case, for $z_1=1,z_2=e^{2\beta}$ and $v=1$.\\
\\
{\bf Figure 5:} (Color online) Shows the free energy density as a function of the average
density in the $a=2,b=0$ case, for $z_1=1,z_2=e^{2\beta}$ and $v=1$.\\
\\
{\bf Figure 6:} (Color online) The upper panel shows the RPA validity region of the phase
diagram, in the Janus case, for
$z_1=1,z_{10}=e^{10\beta},z_{40}=e^{40\beta}$, in the 
absence of any excluded volume effect. At
$\beta=0$ the $C=0$ equation has solution $\rho\simeq
0.647933\ldots$. Note that the validity region is in the small density
region, where the contribution from the quadratic fluctuations of the
theory dominates, and the whole theory is expected to be less
significant. The lower panel shows the free energy density as a
function of the average density. The rapid increase at
high density is indicative of the limit of the RPA applicability.\\
\\
{\bf Figure 7:} (Color online) Shows the free energy density as a function of the average
density in the Janus case, for
$z_1=1,z_{10}=e^{10\beta},z_{40}=e^{40\beta}$, and $v=15$.\\
\\
{\bf Figure 8:} (Color online) In the Janus case for
$z_1=1,z_{10}=e^{10\beta},z_{40}=e^{40\beta}$, $v=15$, and $\beta=1$ shows the
concentrations of clusters of 1, 10, and 40 particles as a function of
the density.
\\
{\bf Figure 9:} (Color online) In the Janus case for
$z_1=1,z_{10}=e^{10\beta},z_{40}=e^{40\beta}$, and $v=15$, shows the
concentrations of clusters of 1 ($n_1^S$), 10 ($n_{10}^S$), and 40 ($n_{40}^S$)
particles as a function of the density when we do not use the
logarithmic correction in Eq. (\ref{M}).
\end{document}